\documentclass[a4paper,fleqn,usenatbib]{mnras}
  
\usepackage{newtxtext,newtxmath}

\usepackage[T1]{fontenc}
\usepackage{ae,aecompl}

\usepackage{graphicx}	
\usepackage{amsmath}	
\usepackage{amssymb}	

\def\be{\begin{equation}}
\def\ee{\end{equation}}
\usepackage{rotating}
\usepackage{multirow}

\title[Bulk flow in 2MTF and 6dFGSv]{Bulk flow in the combined 2MTF and 6dFGSv surveys}

\author[F. Qin et al.]{
Fei Qin$^{1,2}$\thanks{E-mail: fei.qin@research.uwa.edu.au},
Cullan Howlett$^{1,2}$,
Lister Staveley-Smith$^{1,2}$,
Tao Hong$^{3,2}$
\\
$^{1}$International Centre for Radio Astronomy Research, University of Western Australia, 35 Stirling Hwy, Crawley, WA 6009, Australia\\
$^{2}$ARC Centre of Excellence for All-sky Astrophysics (CAASTRO)\\
$^{3}$National Astronomical Observatories, Chinese Academy of Sciences, 20A Datun Road, Chaoyang District, Beijing 100101, China
}

\date{Accepted XXX. Received YYY; in original form ZZZ}

\pubyear{2017}

\begin{document}
\label{firstpage}
\pagerange{\pageref{firstpage}--\pageref{lastpage}}
\maketitle

\begin{abstract}
We create a combined sample of 10,904 late and early-type galaxies from the 2MTF and 6dFGSv surveys in order to accurately measure bulk flow in the local Universe. Galaxies and groups of galaxies common between the two surveys are used to verify that the difference in zero-points is $<0.02$ dex.
We introduce a maximum likelihood estimator ($\eta$MLE) for bulk flow measurements which allows for more accurate measurement in the presence non-Gaussian measurement errors. To calibrate out residual biases due to the subtle interaction of selection effects, Malmquist bias and anisotropic sky distribution, the estimator is tested on mock catalogues generated from 16 independent large-scale GiggleZ and SURFS simulations. The bulk flow of the local Universe using the combined data set, corresponding to a scale size of 40 h$^{-1}$ Mpc, is $288\pm24$ km s$^{-1}$ in the direction $(l,b)=(296\pm6^{\circ}, 21\pm5^{\circ})$. This is the most accurate bulk flow measurement to date, and the amplitude of the flow is consistent with the $\Lambda$CDM expectation for similar size scales.
\end{abstract}

\begin{keywords}
galaxies: kinematics and dynamics-galaxies: statistics-large-scale structure of
the Universe-surveys.
\end{keywords}



\section{Introduction}

The Universe is expanding, driving galaxies further and further apart from each other. Locally, this  effect is described by the Hubble Law, which linearly relates the distance and redshift of galaxies. However, on top of the `cosmological expansion' or `Hubble Flow', additional velocity components may be present. A major additional component is the result of the combined gravitational effects of mass density fluctuations. These give rise to perturbations on the Hubble flow, called `peculiar velocities', or just `velocities'. These peculiar velocities are good indicators of the density field in the nearby Universe and enable us to determine cosmological parameters, test the cosmological model, and test whether General Relativity accurately describes the motion of galaxies on the largest scales, given the observed density field.

Coherent peculiar velocities over a significant volume of the Universe are known as `bulk flows'. These flows seem to arise from massive (supercluster-scale) overdensities, and can therefore provide a measurement of the total mass of these overdensities, and correspondingly constrain the degree of homogeneity and isotropy required in cosmological models. Bulk flow in the local Universe is simply defined as the weighted mean value of the line-of-sight peculiar velocities, projected on three orthogonal axes, for a given galaxy sample. It is an important measurement of the density field on scales comparable or greater than the sampled volume. However, measuring the bulk flow is prone to systematic effects arising from poor distance estimation and sample selection. This is because measurements of both redshift and a redshift-independent distance are required. 

An early claim of bulk flow in the local Universe was made by \cite{1976AJ.....81..719R}, who found a bulk flow velocity of 534 km~s$^{-1}$ under the direction $l=73^{\circ}$, $b=19^{\circ}$ (Galactic coordinates) in the CMB frame (or 600 km~s$^{-1}$ in the direction $(l,b)=(160^{\circ},-10^{\circ})$ of the local group (LG) frame). 
Later measurements, for example by \cite{1988ApJ...326...19L}, converged on a different bulk flow direction $l=312\pm11^{\circ}$, $b=6\pm10^{\circ}$, showing the earlier direction to be erroneous. However, subsequent studies have also shown that the amplitude is likely not as high as 600 km~s$^{-1}$. \cite{1993ApJ...412L..51C} estimated $360\pm40$ km~s$^{-1}$ in the direction $l\approx300^{\circ}$, $b\approx10^{\circ}$. \cite{2009MNRAS.392..743W} found $407\pm81$ km~s$^{-1}$, with a direction of $l=287\pm9^{\circ}$, $b=8\pm6^{\circ}$. More recently, for the 2MTF survey \cite{2014MNRAS.445..402H} found $281\pm 25$ km~s$^{-1}$ in the direction of $l=296\pm16^{\circ}$, $b=19\pm6^{\circ}$. With the deeper 6dFGSv survey, \cite{2016MNRAS.455..386S} derived a bulk flow of $248\pm 58$ km~s$^{-1}$ in the direction $(l,b) \approx (318^{\circ},40^{\circ})$. 

Quantitative comparison between the above results, and comparison with the predictions of $\Lambda$CDM cosmology, needs to take into account the window function which describes the effective depth of each survey, and is fairly straightforward to calculate. But the comparison, and indeed the window function itself, also depends on estimation technique. In recent literature, the main techniques that have been used to calculate bulk flow are: maximum likelihood estimation (MLE) \citep{1988MNRAS.231..149K,2007MNRAS.375..691S}; minimum variance (MV) estimation \citep{2009MNRAS.392..743W,2010MNRAS.407.2328F} and log-linear
$\chi^2$ minimization \citep{2014MNRAS.445..402H}. MLE is fast and easy to use, but incorrectly assumes that peculiar velocity measurements have a Gaussian distribution. MV takes into account large-scale correlations, but is slow, not ideal for large data sets, and still assumes Gaussian errors. \cite{2015MNRAS.450.1868W} propose a velocity estimator that allows either of the above methods to be used with less bias.

In this paper, we combine the deep, but hemispherical 6dFGSv data \citep{2014MNRAS.445.2677S} with the more isotropic but shallower 2MTF data \citep{2014MNRAS.445..402H} in order to better understand and measure velocities and bulk flows in the local Universe. Both surveys have well-defined selection functions and we are able to model the observational selection effects by using mock surveys generated from large-scale simulations. Moreover, we have also investigated different velocity estimators, and different bulk flow estimates, and we propose a log-linear MLE technique based on magnitude fluctuations -- the so-called $\eta$MLE method.  

This paper is structured as follows: in Section 2, we give a brief introduction to 2MTF, 6dFGSv and the group catalogue which we use for comparison. In Section 3, we discuss the combination of 2MTF and 6dFGSv. In Section 4, we introduce the peculiar velocity estimators, including a $\eta$MLE algorithm. In Section 5, we discuss results obtained from the mocks. We present our final results in Section 6 and conclude in Section 7.

For analysing the data, we adopt a spatially flat $\Lambda$CDM cosmology is as the fiducial model. The cosmological parameters are from Planck Collaboration (2013): $\Omega_m=0.3175$, $\sigma_8=0.8344$, and $H_{0} = 100 h$ km s$^{-1}$ Mpc$^{-1}$ with $h=0.67$ \cite{2014A&A...571A..16P}. These parameters are mainly used to calculate the comoving distance and the power spectrum, and don't significantly affect the results.

\section{DATASETS} \label{sec:data}

\subsection{2MTF}
 
 The Two Micron All-Sky Survey (2MASS) Tully-Fisher
Survey (2MTF) uses high-quality velocity widths from 21-cm HI observations and photometry from the 2MASS survey to measure Tully-Fisher distances for bright inclined spirals in the 2MASS Redshift Survey (2MRS) \citep{2008AJ....135.1738M,2012ApJS..199...26H}.

All 2MTF photometric data are obtained from the 2MRS catalogue. The HI rotation widths are obtained from archival data \citep{2005ApJS..160..149S}, and from new observations with the GBT \citep{2014MNRAS.443.1044M} and Parkes telescopes \citep{2013MNRAS.432.1178H}, as well as from new ALFALFA data \citep{2011AJ....142..170H}. The final 2MTF catalogue is selected using the following criteria: total $K$-band magnitude $K<11.25$ mag,  600 km~s$^{-1}$ $<cz<$ 10,000 km~s$^{-1}$, axis ratio $b/a<0.5$, HI spectrum signal-to-noise ratio SNR$>5$, HI width error $\epsilon_w / \epsilon_{HI}<10\%$. The final 2MTF sample includes 2062 galaxies, down to the Galactic latitude $|b|=5^{\circ}$.

\subsection{6dFGSv}
The Six-degree-Field Galaxy Survey (6dFGS) is a southern survey, also based on the 2MASS near-infrared galaxy catalogue, and covering about 17,046 deg$^2$, with Galactic latitude $|b|>10^{\circ}$ out to $cz \approx 16,500$ km~s$^{-1}$. The targets are galaxies with total K-band magnitude $K<12.75$ mag in the 2MASS Extended
Source Catalog  \citep{2009MNRAS.399..683J,2004MNRAS.355..747J}. The 6dFGS Catalog contains about 150,000 galaxies.

The 6dFGS peculiar-velocity survey (6dFGSv) contains the brightest early-type galaxies in the primary 6dFGS redshift sample with SNR $>5$ and total $J$-band magnitude $J<13.65$ mag, out to $cz = 16,500$ km~s$^{-1}$, and velocity dispersion greater than 112 km~s$^{-1}$ \citep{2014MNRAS.445.2677S}.
Distances and velocities of the 6dFGSv sample are measured by using the fundamental
plane technique. The
best-fitting fundamental
plane and peculiar velocities have been determined for 8,885 6dFGSv galaxies
in near-infrared passbands \citep{2012MNRAS.427..245M}.   

\subsection{Group Catalog}

 The galaxies group catalogs, which will be used in the following research, is the group list identified by \cite{2007ApJ...655..790C}, including the low density contrast (LDC) catalog, and the hight density contrast (HDC) catalog. The group catalog will be used to verify the zero-points between the 2MTF and the 6dFGSv.

\section{Combining 2MTF and 6dFGSv}

The combined 2MTF and 6dFGSv data offer a number of advantages over each survey individually. Firstly, the 6dFGSv survey, which extends out to $cz\approx16,200$ km~s$^{-1}$, is deeper than 2MTF alone. Secondly, 2MTF is more uniform than 6dFGSv alone, and provides vital coverage in the northern hemisphere. Finally, the combination of the data sets will statistically improve the estimate of bulk flow and other cosmological parameters. In related work, \cite{2017MNRAS.464.2517H} has predicted that such a data set combination can improve the measurement of the normalised growth factor, $f\sigma_{8}$ by 25\%. For consistency, we use $J$-band 2MASS photometry throughout.
 
Before combining 2MTF and 6dFGSv, we need to cross-compare the distance estimates and correct for any the zero-point offsets. Such offsets may be due to the systematic differences between the Tully-Fisher and Fundamental Plane distances, different measurement techniques, or different selection effects. We define the logarithmic distance ratio for a galaxy as:
\be\label{logd}
\eta \equiv \log_{10}\frac{d_z}{d_h}
\ee
where $d_z$ is the apparent distance of a galaxy as judged from its redshift, and $d_h$ is the true comoving distance of the galaxies inferred from the Fundamental Plane or Tully-Fisher relation. The observed redshift includes the line-of-sight peculiar motion and the cosmological expansion.

\subsection{Common galaxies}

43 2MTF galaxies have 2MASS IDs listed in Table 2 of \cite{2014MNRAS.443.1231C}. These galaxies all have 2MTF and 6dFGSv velocity differences c$|\Delta z| < 150$ km s$^{-1}$. The properties of these galaxies are listed in Table~\ref{tbbb}. Most appear to be late-type galaxies misclassified in 6dFGSv due to having an early-type nuclear spectrum, as shown in Figure 15 in \cite{2014MNRAS.443.1231C}.

\begin{figure}  
 \includegraphics[width=\columnwidth]{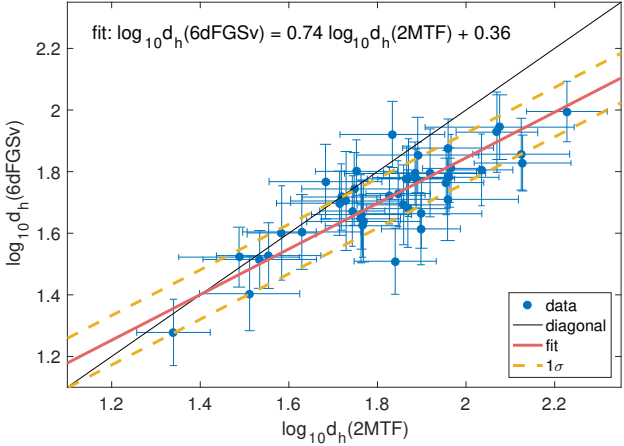}
 \caption{A comparison of 2MTF and 6dFGSv distances for 43 common galaxies. The solid red line is the Hyper Fit line, and the yellow dashed lines represent $\pm1\sigma$, where $\sigma=0.08$. The solid black line is the expected 1:1 relation for perfect distance estimators.}
\label{commongals}
\end{figure}

In Fig.\ref{commongals}, we plot $\log_{10}d_h$(2MTF) against $\log_{10}d_h$(6dFGSv) for the above common galaxies, and use the Hyper Fit package \citep{2015PASA...32...33R} to perform a weighted fit. The average difference for the common galaxies is
\be
\Big\langle\log_{10}\frac{d_h(2MTF)}{d_h(6dFGSv)}\Big\rangle=0.11\pm0.01
\ee
representing a zero-point offset in distance of almost 30\%. The Fundamental Plane distances appear abnormally small due to the late-type nature of the galaxies.

\subsection{Common groups}

A better method to study the zero-point offset is to compare distance estimates of galaxies in the same groups, and therefore at common distances. In that way, misclassified galaxies are avoided. We pick common groups using the low-density-contrast (LDC) catalogue of \cite{2007ApJ...655..790C}.

We firstly remove the 43 common galaxies, then identify the group IDs for the 2MTF galaxies and the 6dFGSv galaxies. After galaxies are assigned to a group ID, we pick out those groups which contain both 2MTF and 6dFGSv galaxies. We find 95 LDC common groups.   

\begin{figure}  
 \includegraphics[width=\columnwidth]{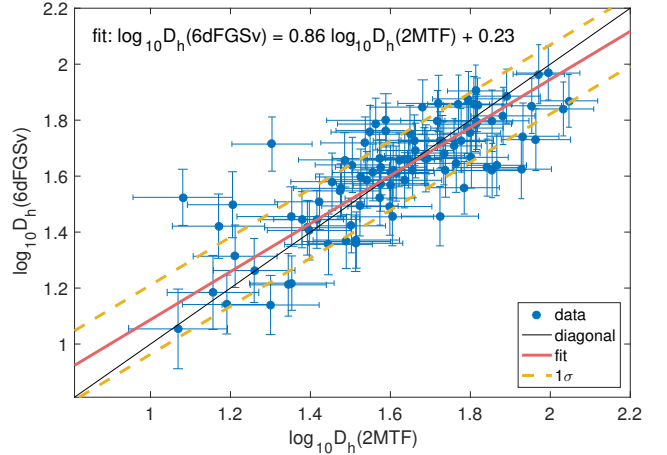}
 \caption{A comparison of 2MTF and 6dFGSv distances for 94 common groups (group IDs are identified by using the LDC). The solid red line is the Hyper Fit line, and the yellow dashed lines represent $\pm1\sigma$, where $\sigma=0.12$. The solid black line is the expected 1:1 relation for perfect distance estimators.}
\label{commongro}
\end{figure}

For each group, we calculate the mean $\eta$ and the mean $\log_{10}d_z$ of the 2MTF and 6dFGSv galaxies. As a result, each group has a 2MTF distance:
 \be
 \log_{10}D_{h}(2MTF)= \langle   \eta(2MTF)  \rangle - \langle \log_{10}d_z(2MTF) \rangle
 \ee
 and a 6dFGSv distance:
  \be
 \log_{10}D_{h}(6dFGSv)=\langle   \eta(6dFGSv)  \rangle - \langle \log_{10}d_z(6dFGSv) \rangle .
 \ee 
A linear fit followed by a $3\sigma$ clip removes one LDC group, leaving 94. In Fig.~\ref{commongro}, we plot the LDC identified $\log_{10}D_{h}(2MTF)$ against $\log_{10}D_{h}(6dFGSv)$ and obtain a regression line much closer to the expected diagonal. 
We obtain the average value for the logarithmic distance ratio of 
\be\label{ssssaa}
\Big\langle \log_{10}\frac{D_{h}(2MTF)}{D_{h}(6dFGSv)} \Big\rangle=0.00\pm0.02 
\ee
which is consistent with zero.

From the 51 common groups (again, after a 3$\sigma$ clip) in the HDC catalogue of \cite{2007ApJ...655..790C}, the average value for the logarithmic distance ratio is $0.00\pm0.02$, also consistent with zero.

\begin{figure*} 
 \includegraphics[width=175mm]{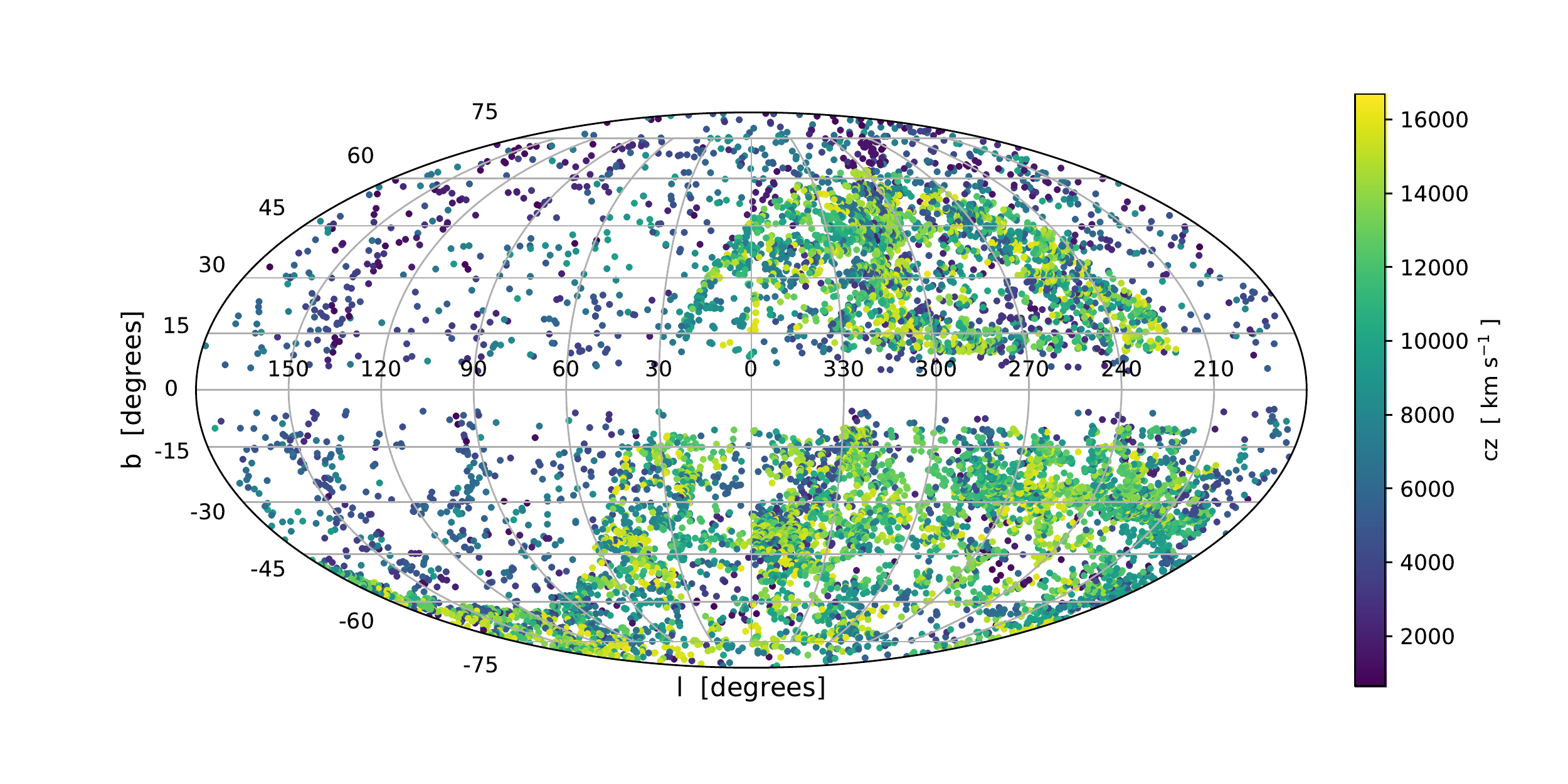}
 \caption{Sky coverage in galactic coordinates of 10,904 galaxies in the combined data sets of 2MTF and 6dFGSv. The colour of the points refers to the galaxy redshift, according to the colour bar on the right-hand side of the plot.}
 \label{lb}
\end{figure*}

\begin{figure} 
\centering
 \includegraphics[width=\columnwidth]{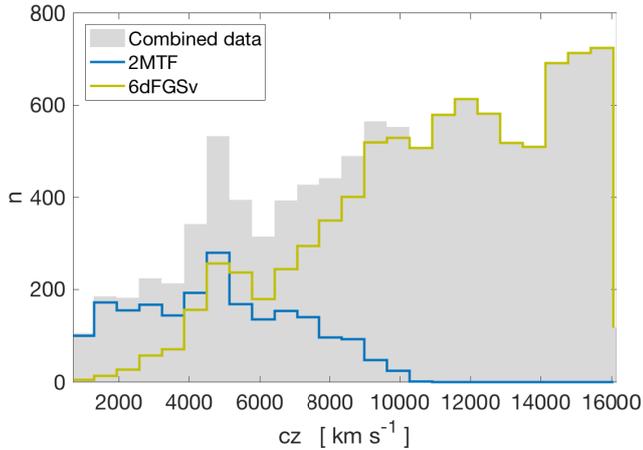}
 \caption{Redshift distribution of the data sets in the CMB frame. The gray bars are for the combined data set. The light-green line is for the 6dFGSv, while the blue line is for the 2MTF.}
 \label{histcz}
\end{figure}

\subsection{The combined data set of the 2MTF and the 6dFGSv}

No significant zero-point correction is required, so we simply remove the 43 common galaxies from 6dFGSv and combine the remaining galaxies with 2MTF, resulting in a combined data set of 10,904 galaxies. The sky coverage of the combined data set is shown in Fig.~\ref{lb} and the redshift distribution in Fig.~\ref{histcz}.

\section{Maximum Likelihood Bulk Flow Estimation} \label{sec:theory}

We will use the measurements of the logarithmic distance ratio, $\eta$, or their corresponding peculiar velocities, $v$ from the combined 2MTF and 6dFGSv samples to estimate the bulk flow of our local Universe. For the $\Lambda$CDM model, we expect the observed set of large scale velocities to be drawn from a Gaussian distribution with variance related to the velocity power spectrum smoothed over some characteristic scale and mean equal to the observers own bulk flow. As the distribution of velocities (at fixed characteristic scale or depth) is based on the velocity power spectrum, so too is the distribution of possible bulk flows. Hence, measurements of our local bulk flow at different depths can be used to test the cosmological model. If such measurements agree with the distribution of possible $\Lambda$CDM predicted bulk flows, then $\Lambda$CDM constitutes a probable cosmological model. If we were to measure a bulk flow significantly outside the expected distribution of bulk flows this could indicate that the $\Lambda$CDM model is incorrect.
 
Measuring the bulk flow velocity at some depth requires finding the value for the 3-dimensional velocity at the location of an observer that maximises the likelihood of observing a particular configuration of log-distance ratios or peculiar velocities. However, this process is complicated by the fact that the relationship between $\eta$ and $v$ is non-linear and that real observations of galaxy velocities are only along the line-of-sight and often have some degree of anisotropy and (possibly non-Gaussian) measurement error. With these assumptions and caveats in mind, in this section we present three methods for measuring the maximum likelihood bulk flow given a set of $\eta$ measurements, which will then be tested using our mock galaxy catalogues in Section~\ref{sec:mocktests}.

All three of these methods are based on similar maximum likelihood methods, however they differ crucially in how they model the relationship between the measurements of $\eta$ and the model bulk flow, $\vec{B}$. The first two convert the measurements of $\eta$ into peculiar velocities using different estimators, namely those of \cite{2014MNRAS.442.1117D} and \cite{2015MNRAS.450.1868W}. As such we name these the \textit{d}MLE and \textit{w}MLE bulk flow estimators. Our third method, which we will show is superior for the 2MTF and 6dFGSv datasets, instead converts the model bulk flow from velocity-space to $\eta$-space and compares this directly to the measurements. We name this the $\eta$MLE method.

\subsection{Maximum likelihood bulk flow in velocity space} \label{sec:MLE}
\cite{1988MNRAS.231..149K} writes the likelihood of observing a set of $n$ peculiar velocities $v_{n}$ given a bulk flow $\vec{B}$ as
\be\label{tramle}
L(\vec{B},\sigma_{\star})=\prod_{n=1}^{N}\frac{1}{\sqrt{2 \pi \left(\sigma^2_n+\sigma_{\star}^2 \right)}}\exp\left(-\frac{1}{2}  \frac{    (v_{n}-\vec{B}\cdot \hat{r}_n)^2  }{  \sigma^2_n+\sigma_{\star}^2 }\right)
\ee
where $ \hat{r}_n$ is the unit vector pointing to the $n$-th galaxy, $\sigma_n$ is the measurement error of $v_{n}$ and $\sigma_{\star}$ is the typical 1D non-linear velocity dispersion, usually assumed to be $\sim 300$ km s$^{-1}$ \citep{2007MNRAS.375..691S,2016MNRAS.455..386S}. The maximum likelihood value of the bulk flow vector can then be obtained by writing each component as the weighted mean value of the line-of-sight peculiar velocities,
 \be\label{bkf}
B_i\equiv  \sum^N_{n=1}w_{n,i}v_{n}~,~~(i=1,2,3),
\ee
and maximizing $L(\vec{B},\sigma_{\star})$. Doing so shows that the weight factors can be analytically expressed as 
 \be \label{aijml3}
w_{n,i}=\sum^3_{j=1}A_{ij}^{-1}\frac{\hat{r}_{n,j}}{\sigma^2_n+\sigma^2_{\star}}~,~~(i=1,2,3)
\ee
where 
\be\label{aijml}
A_{ij}=\sum^N_{n=1}\frac{\hat{r}_{n,i}\hat{r}_{n,j}}{\sigma^2_n+\sigma^2_{\star}} .
\ee
The `MLE depth' of the bulk flow measurement, i.e., the characteristic scale at which the measurement of the bulk flow should be compared to the theoretical expectation, is similarly defined via
\be\label{depthri}
d_{MLE}=\frac{\sum |\boldsymbol{r}_n|W_n}{\sum W_n}
\ee
where we now use weights $W_n=1/(\sigma_n^2+\sigma^2_{\star})$.

The measurement error of the bulk flow consists of two parts \citep{1988MNRAS.231..149K,2010MNRAS.407.2328F}:
\be\label{rijl}
R_{ij}= \langle  B_iB_j  \rangle =R^{B}_{ij}+R^{\epsilon}_{ij}.
\ee
$R^{B}_{ij}$ is the cosmic variance term, which arises from the finite volume in which the bulk flow is measured and is intrinsically linked to the velocity power spectrum and the characteristic depth of the bulk flow measurement. This will be revisited in Section~\ref{sec:theorycomp}. $R^{\epsilon}_{ij}=A^{-1}_{ij}$ is the measurement covariance matrix for each of the bulk flow components, where the equality results from the maximum likelihood solution to the Gaussian likelihood in Eq.~\ref{tramle}.
Finally, the variance of the bulk flow \textit{amplitude} is \citep{2016MNRAS.455..386S}:
\be\label{bke2}
e^2_B=JR^{\epsilon}_{ij}J^T~,~~(i=1,2,3)
\ee
where $J$ is the Jacobian of the bulk flow, $\partial B/\partial B_i$.

\subsubsection{\textit{d}MLE} \label{sec:dMLE}

Neglecting the effects of relativistic motions and gravitational lensing, the line-of-sight velocity of a galaxy is related to its measured redshift, $z$, using \citep{2014MNRAS.442.1117D},
\be \label{travp}
v=\boldsymbol{v}\cdot{\boldsymbol{\hat{r}}}=c\left(\frac{z-z_h}{1+z_h}\right)
\ee
where $c$ is the speed of light and $z_h$ is the redshift corresponding to the galaxy's true comoving distance $d_h$. In the flat $\Lambda$CDM model
\be\label{Dz}
d_{h}(z_{h})=\frac{c}{H_0}\int_0^{z_{h}}\frac{dz'}{E(z')}\approx \frac{cz_{h}}{H_{0}}
\ee
where
\be\label{Ez}
E(z)=\frac{H(z)}{H_0}=\sqrt{\Omega_m(1+z)^3+\Omega_{\Lambda}},
\ee
and $H(z)$ is the Hubble constant corresponding to redshift $z$, which at the present epoch is given by $H_{0}$, and $\Omega_{m}$ and $\Omega_{\Lambda}$ are the present day matter and dark energy densities, respectively. A similar expression relates the observed redshift $z$ to the \textit{inferred} comoving distance $d_z$ for a given cosmological model. At low redshifts we can approximate $d_{h}(z_{h}) \approx cz_{h}/H_{0}$.

From measurements of $\eta$ and the inferred comoving distance $d_{z}$, we can compute the peculiar velocity by first calculating
\be \label{etadist}
d_{h} = d_{z}10^{-\eta}
\ee
then solving Eq.~\ref{Dz} and Eq.~\ref{travp} for $z_{h}$ and $v$ respectively. However, measurements of the Tully-Fisher and Fundamental Plane relations typically return Gaussian errors in the log-distance ratio $\eta$ which translate to log-normal distributed errors in the velocity, even in the low redshift approximation.

Hence, a careful choice of how to extract a measurement and error on the peculiar velocity given a mean and error on $\eta$ must be made. \cite{2016MNRAS.455..386S} demonstrate that the mean value of the $\eta$ distribution is equivalent to the median value of the corresponding $v$ distribution and that the standard deviation of the velocity distribution has a strong linear dependence on $d_{z}$ that can be modelled as,
\be\label{dcz1}
\sigma_{n}(6dFGSv)=0.324H_0d_{z,n}.
\ee
for the 6dFGSv dataset. Applying the same procedure to the 2MTF data (Appendix~\ref{AP1}) we find
\be\label{dcz2}
\sigma_{n}(2MTF)=0.177H_0d_{z,n}.
\ee

Based on the above relationship between $\eta$ and $v$, the \textit{d}MLE method uses the peculiar velocity based on the mean log-distance ratio and the error on $v$ given by Eq.~\ref{dcz1} and Eq.~\ref{dcz2} for the 6dFGSv and 2MTF data as input to the maximum likelihood method. Because this method of estimating the peculiar velocity and bulk flow does not fully encapsulate the error distribution for each galaxy, we expect (and show in Section~\ref{sec:mocktests}) that this method is suboptimal for measuring the bulk flow in the combined 2MTF and 6dFGSv data.

\subsubsection{wMLE} 

An alternative method of estimating the peculiar velocity given a measurement of the $\eta$, which largely preserves the Gaussian nature of the error distribution, is given by \cite{2015MNRAS.450.1868W},
\be\label{watvp}
v=\frac{\ln(10)cz}{1+z}\log_{10}\frac{cz}{H_0d_h} \approx \frac{\ln(10)cz}{1+z}\eta,
\ee
where the second equality arises from the low redshift approximation of the log-distance ratio. A similar expression can then be used to calculate the error on the peculiar velocity given the error on $\eta$. We call the combination of this peculiar velocity estimator and the likelihood in section~\ref{sec:MLE} the $w$MLE method.

One caveat to this is that it only strictly returns an unbiased estimate of a peculiar velocity under the assumption that the galaxy's \textit{true} peculiar velocity (not necessarily the measured peculiar velocity which can be significantly larger) is much smaller than $cz$ for that galaxy. Typical galaxy velocities are expected to be on the order of a $500-1000$ km s$^{-1}$, hence at first glance this is a reasonable approximation for most of the galaxies in the 2MTF and 6dFGSv samples which have mean redshifts of $\sim 5000$ km s$^{-1}$ and $\sim 10000$ km s$^{-1}$ respectively. However, we note that this approximation results from a Taylor expansion of $\mathrm{ln}(1-v/cz)$ which typically converges poorly and that the most accurate distances measurements in the two surveys are at low redshift where the approximation is more likely to break down. As a result of this, although the estimator is found to perform reasonably well, in the next section we also develop an estimator which does not require this assumption.
 
\subsection{$\eta$MLE: Estimation in log-distance space}\label{newmleee}

In this section we introduce a algorithm for estimating the bulk flow that preserves the Gaussian nature of the measurement errors and that does not require any assumption on the unknown magnitude of a galaxy's velocity compared to its redshift. The principle behind this estimator is that rather than converting the measurements of $\eta$ to velocities and using these as input to the \cite{1988MNRAS.231..149K} maximum likelihood method, we take a more Bayesian approach; calculating theoretical log-distance ratios for each galaxy given a model for the bulk-flow, then comparing these directly to the measurements.

Starting with the assumption that the measured log-distance ratios for a given set of galaxies are independent and Gaussian distributed, we can write the likelihood of observing a particular set of log-distance ratios as 
\be\label{pvpi}
P(\boldsymbol{\eta} | \vec{B})=\prod^{n}_{i=1}\frac{1}{\sqrt{2 \pi \left(\epsilon_{i}^2+\epsilon_{\star,i}^2 \right)}}\exp\left({-\frac{1}{2}  \frac{    (\tilde{\eta_{i}}(\vec{B})-\eta_{i})^2  }{  \epsilon_{i}^2+\epsilon_{\star,i}^2 }}\right),
\ee
where $\epsilon_{i}$ is the measurement error of $\eta_{i}$ for each galaxy and $\epsilon_{\star,i}$ encapsulates the effects of non-linear motions on the measurements. We relate $\epsilon_{\star,i}$ to the usual non-linear parameter $\sigma_{\star}$ using \citep{2014MNRAS.444.3926J,2017MNRAS.471.3135H}
\be 
\epsilon_{\star,i}= \frac{1+z_{i}}{\mathrm{ln}(10)H(z_{i})d_{z,i}}\sigma_{\star}. 
\ee
The above equation results from the derivation of \cite{2006PhRvD..73l3526H} which demonstrates how a peculiar velocity changes the observed magnitude of a galaxy, which is in turn related to the log-distance ratio. This expression technically involves a similar Taylor expansion to that of \cite{2015MNRAS.450.1868W}, but as we treat $\sigma_{\star}$ as a free, nuisance parameter this approximation is expected to be much less important.

$\tilde{\eta_{i}}(\vec{B})$ is the log-distance ratio that each observed galaxy would have if its velocity was equal to a bulk flow $\vec{B}$. The procedure to calculate this can be inferred from section~\ref{sec:dMLE}:
\begin{enumerate}
\item{Calculate the line-of-sight velocity $v(\vec{B})=\vec{B} \cdot \hat{r}$ for a galaxy at position $\boldsymbol{r}$ due to a bulk flow velocity $\vec{B}=\{B_x,B_y,B_z\}$.}
\item{Evaluate the predicted true comoving distance to the galaxy, $d_{h}$, based on the observed redshift and Eqs.~\ref{travp} and \ref{Dz}.}
\item{Calculate the model $\tilde{\eta_{i}}(\vec{B})$ given the known (for a given cosmological model) $d_{z}$ and predicted $d_{h}$}.
\end{enumerate}

The non-linear transformation of the model $\vec{B}$ to a predicted log-distance ratio for each galaxy means that the maximum likelihood bulk flow cannot be obtained analytically. Instead we combine the likelihood in Eq.~\ref{pvpi} with uniform priors on the bulk flow components and $\sigma_{\star}$, which allows us to write the posterior distribution of these four parameters given our data and a cosmological model.

A similar estimator for the velocity field which forward models
the measured log-distance ratios given a model bulk flow was introduced
by \cite{2011ApJ...736...93N}. The main difference lies in
the conversion from a model bulk flow to a log-distance ratio for
each galaxy, and how the best-fit bulk flow is identified \citep{2011MNRAS.413.2906D}.

In this work, we use a standard Metropolis-Hastings Markov
Chain Monte Carlo (MCMC) algorithm and priors of a flat distribution in the interval $B_{i}\in[-800,+800]$ km s$^{-1}$ to explore the posterior. Even though each likelihood evaluation requires computing predicted log-distance ratios for each galaxy, the above steps are simple enough that obtaining converged posterior distributions is not computationally intensive.
 
\section{Bulk flow fitting for the mocks}\label{sec:mocktests}

To test the three estimators introduced in Section~\ref{sec:theory} and explore how well we expect them to recover the true bulk flow of the combined 2MTF and 6dFGSV data we created a set of realistic mock galaxy catalogues that match the selection function and the survey geometry of these two surveys.

In total, we created $2\times8$ 2MTF mocks and $2\times8$ 6dFGSv mocks based on the GiggleZ \citep{2015MNRAS.449.1454P} and the SURFS simulations \citep{2018MNRAS.475.5338E}. The GiggleZ simulation is $1\,h^{-3}\mathrm{Gpc}^{3}$ in size, has a halo mass resolution of $3.0\times10^{11}h^{-1}\,M_{\odot}$ and uses a WMAP-5 cosmology ($\Omega_m=0.273$, $\Omega_b=0.0456$, $\sigma_8=0.812$, and $h=0.705$). The SURFS simulation is slightly smaller at $900^{3}\,h^{-3}\mathrm{Mpc}^{3}$ and uses a \textit{Planck}-based cosmology ($\Omega_m=0.3121$, $\Omega_b=0.0488$, $\sigma_8=0.815$, and $h=0.6751$), but with a similar halo mass limit of $1.5\times10^{11}h^{-1}\,M_{\odot}$. Using two different simulations allows us to create a larger sample of independent mocks and to ensure that the estimators of the bulk flow give consistent answers for different cosmologies.

The method for reproducing the 2MTF and 6dFGSv selection functions is quite different, however in both cases galaxies are placed into halos using Subhalo Abundance Matching (SHAM; \citealt{2006ApJ...647..201C}). The exact method for producing these is given below. Each pair of 2MTF and 6dFGSv mocks are created using the same observers, i.e., placing the origin of each pair of mock surveys at the same location, so that they can be combined easily and in the same way as the real data. Based on the number of mock surveys created from the two simulations, each of our combined mocks is non-overlapping and so we treat these as 16 independent samples in the following.

\subsection{2MTF Mocks}

Our mock 2MTF surveys are created using the same method as \cite{2017MNRAS.471.3135H}. K-band luminosities are drawn from the \cite{2001ApJ...560..566K} fit to the luminosity function and are assigned to each halo and subhalo based on their maximum circular velocity. The position and velocity of each halo/subhalo is taken as the position and velocity of the mock galaxy.

From this sample of galaxy positions, velocities and absolute K-band luminosities we reproduce the 2MTF selection function for 8 different observers by applying cuts in redshift of $600\mathrm{\,km~s^{-1}} \le cz \le 10,000\mathrm{\,km~s^{-1}}$ and in \textit{apparent} magnitude of $K < 11.25$ mag. The survey mask is reproduced by removing mock galaxies with galactic latitude $|b|<5^{\circ}$ and down-sampling galaxies to match the redshift distribution of the 2MTF data. This is done separately above and below a declination $\delta=-40.0^{\circ}$ as the number of objects in the true 2MTF dataset is $2.04$ times less below this declination due to the different telescopes used to make the observations.

\begin{figure}  
 \includegraphics[width=\columnwidth]{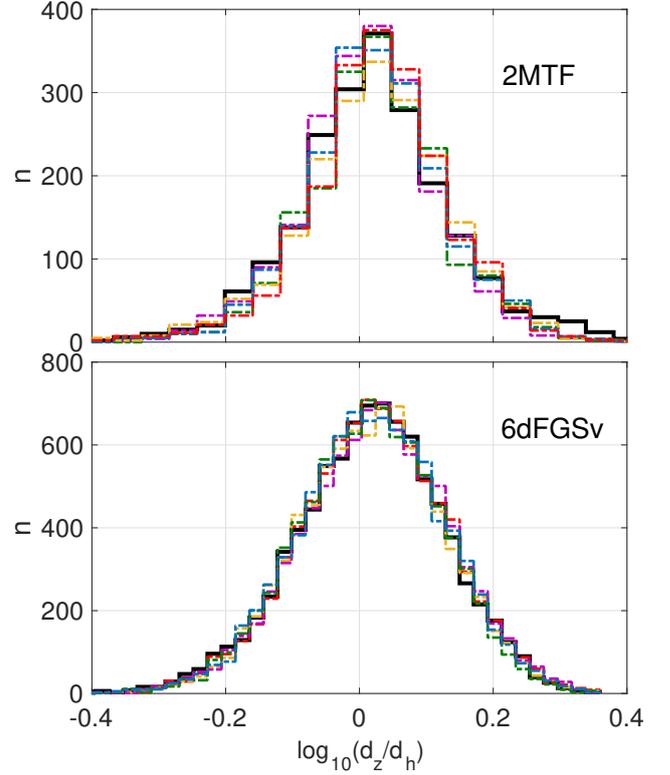}
 \caption{The distribution of $\eta$ for the mocks and the 2MTF and 6dFGSv data. The upper panel is for 2MTF, with the black solid line representing the data, and the (coloured) dashed lines being the distribution of the mocks. Five example mocks are shown. The bottom panel is for 6dFGSv, with the black solid line representing the data, and the (coloured) dashed lines being the distribution of the mocks. Five example mocks are shown.}
\label{logdmock}
\end{figure}

After this process we are left with 16 mock 2MTF catalogues each containing $~\sim 2000$ galaxies. The true velocity of each galaxy is known from the simulation as is the true log-distance ratio. We class the `true' bulk flow vector within each mock as the average of the true galaxy velocities in each direction. Measured log-distance ratios are then calculated for each mock galaxy by drawing from a Gaussian distribution based on the true log-distance ratio with standard deviation given by the fit to the error in the 2MTF measurements as a function of redshift from \cite{2017MNRAS.471.3135H} (Section 2.3.3 therein). In Fig.~\ref{logdmock}, we plot the distribution of the $\eta$ for the 2MTF data, and five example 2MTF mocks. The measured log-distance ratio and errors for each mock catalogue are used as inputs to the three different MLE bulk flow estimators. In Appendix \ref{AP34s} (Fig.~\ref{ssD1} and \ref{ssD2}), we plot the redshift distribution and the sky coverage for the 2MTF data and example 2MTF mocks to show these can faithfully represent the survey geometry of the 2MTF data.

\begin{figure}  
  \includegraphics[width=\columnwidth]{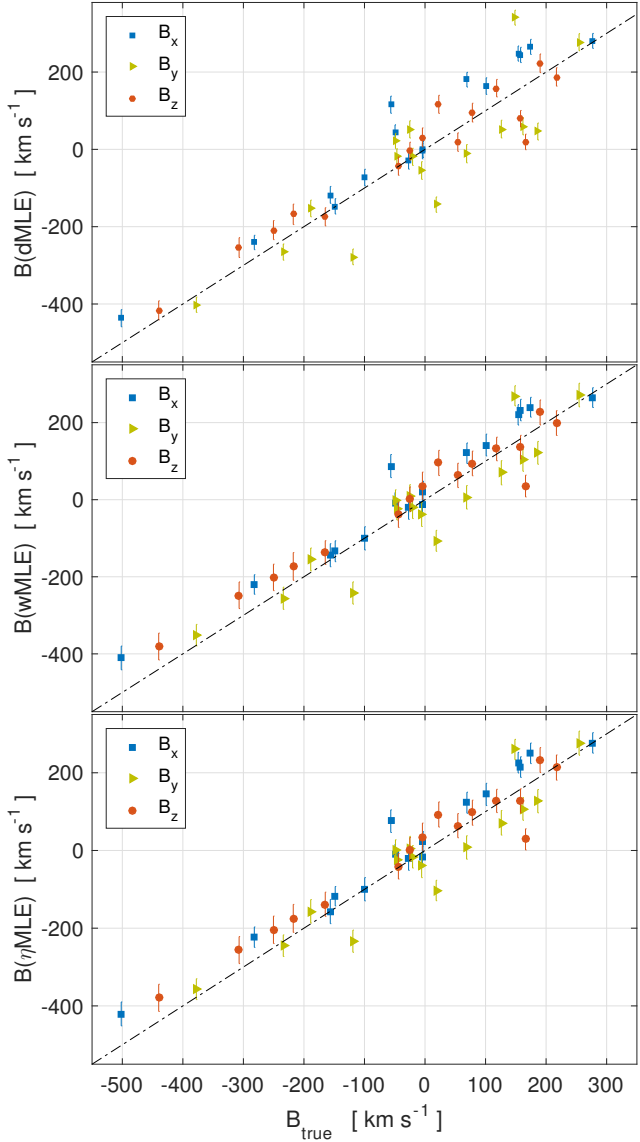}
 \caption{The bulk flow measurement compared to the true bulk flow for the 2MTF mocks in Cartesian equatorial coordinates. The upper panel is for the $d$MLE estimator; the middle panel is for the $w$MLE estimator; the bottom panel is for the $\eta$MLE estimator.}
\label{2mtfmock}
\end{figure}

The resultant bulk flow measurements from the 16 2MTF mocks are plotted against the true bulk flow in Fig.~\ref{2mtfmock}. To compare the three estimators, we calculate the reduced $\chi^2$ between the measured $B_{m}$ and true bulk flow, $B_{true}$ along each direction using
\be\label{chi2}
\chi^2_{red}=\frac{1}{48-1}(\boldsymbol{B}_m-\boldsymbol{B}_{true})\boldsymbol{\mathsf{C}}^{-1}(\boldsymbol{B}_m-\boldsymbol{B}_{true})^T
\ee
where the measured and true bulk flow vectors contain 48 elements (3 directions, and 16 mocks) and $\boldsymbol{\mathsf{C}}$ is the 48$\times$48 covariance matrix. As we treat each of our mocks as independent, this covariance matrix consists of 16 $3\times 3$ sub-covariance matrices on the diagonal and is zero elsewhere. For $w$MLE and $d$MLE, the diagonal blocks of $\boldsymbol{\mathsf{C}}$ are constructed using $R^{\epsilon}_{ij}$ from Eq.~\ref{rijl}, while for the $\eta$MLE the diagonal blocks are calculated by using the 16 MCMC samples. Formulating the reduced chi-squared in this way removes any ambiguity in the number of degrees of freedom and allows us to include the effects of covariance in the three bulk flow components measured in each mock.

For $d$MLE we find $\chi_{red}^2 = 14.45$, which is much larger than for the $w$MLE or $\eta$MLE methods (where $\chi^2_{red}=4.23$ and $4.02$ respectively). This results from an increased scatter in the measured bulk flow about the true bulk flow coupled with the fact that the $d$MLE typically gives the smallest (and least representative) error bars. In general we find that the $w$MLE and $\eta$MLE perform similarly well. The \cite{2015MNRAS.450.1868W} estimator does well in retaining the Gaussian nature of the error bars when transforming from $\eta-$ to $v$-space. However the necessary assumption of $v_{true}\ll cz$ does introduce some systematic error for the closest galaxies in the mocks, and in turn slightly increases the reduced chi-squared compared to the $\eta$MLE. For the 2MTF mocks, we find that the $\eta$MLE is the best performing of the three methods and is the one we adopt for the subsequent parts of this work.

Overall both the $w$MLE and $\eta$MLE have reduced chi-squared values far from unity. Given that both of these methods are expected to account well for measurement errors, the source of this discrepancy comes instead from the underlying assumption of the Maximum likelihood method, that the distribution of true velocities in the mocks is drawn from a Gaussian distribution with mean $\vec{B}\cdot \hat{r}_{n}$ and standard deviation $\sigma_{\star}$ (or equivalently, in the $\eta$MLE method, with mean $\tilde{\eta}_{i}(\vec{B})$ and standard deviation $\epsilon_{\star,i}$). Whilst this assumption is not explicitly stated in the standard maximum likelihood formalism, we can see this from Eq.~\ref{tramle} and~\ref{pvpi}, both of which can be expressed as the convolution of Gaussian distributed true velocities with Gaussian random measurement errors. 

\begin{figure}  
  \includegraphics[height=80mm,width=80mm]{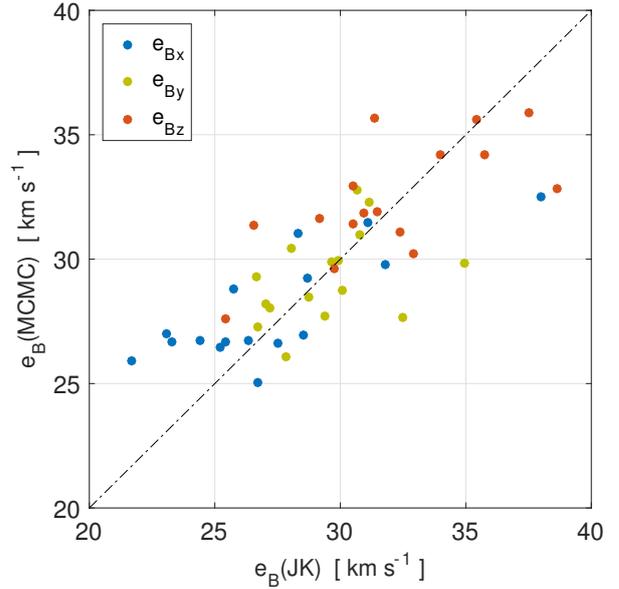}
 \caption{The MCMC errors plotted against the jackknife errors (JK) for the $\eta$MLE bulk flows measured from the 2MTF mocks.}
\label{2mtfmockerrs}
\end{figure}

In reality, non-linearities give rise to large velocities that are not well described by a Gaussian distribution and the above method also does not account for correlations between the velocities of different objects arising from the coherent way in which structures in the Universe grow. This results in systematic modelling errors. One might consider that a way to overcome this is to use jackknife samples to estimate the mean and variance of the measured bulk flow components. As different jackknife samples will be subject to these modelling systematics in different ways, the jackknife error will encapsulate some of the systematic error as statistical error in the measured bulk flow. However, as shown in Fig.~\ref{2mtfmockerrs} for the $\eta$MLE method, we find that the MCMC errors and jackknife errors are largely the same, and the jackknife $\chi^2_{red}=3.71$, which although slightly lower than the MCMC results, does not resolve the discrepancy.

Instead, a more accurate representation of the distribution of velocities is required. This is a significant undertaking and is left for future work. However, we note that the Bayesian approach we adopt with the $\eta$MLE method would allow for rigorous exploration of the effects of different likelihood functions of the bulk flow posterior given a set of measurements.
 
\subsection{6dFGSv Mocks} \label{sec:mocktests6dFGSv}

Having tested our three estimators on the 2MTF mocks and identified that the $\eta$MLE method returns results closest to the true underlying bulk flow, we then turn to mock catalogues based on the 6dFGSv data and the combined sample. We use these to infer how well the $\eta$MLE method is expected to perform on the 6dFGSv data, which contains many more galaxies and pushes to larger depths than the 2MTF data, but with slightly larger log-distance errors and only hemispherical sky coverage.

We created a further $2\times 8$ mocks based on a modified version of the algorithm in \cite{2016MNRAS.455..386S} and \cite{2012MNRAS.427..245M}. Many of the steps follow those presented therein, so will not be reproduced here. Key differences or clarifications in the algorithm we use compared to the above are: 
\begin{enumerate}
\item{Observers in the SURFS and GiggleZ simulations are not chosen at random, but instead are placed at the same locations as the 2MTF mocks.}
\item{Intrinsic (logarithmic) effective radii $r_{t}$, velocity dispersion, $s_t$ and surface brightness, $i_t$ for each mock galaxy are generated from the fundamental plane fit to the $J$-band 6dFGSv data. These quantities are assigned to halos in the simulations by rank-ordering $l=2r_{t}+i_t$ against the halo maximum circular velocity.}
\item{For each galaxy, we compute the apparent size of the galaxy under the influence of its peculiar velocity, which we call $r_{tz}$. Note that this does not include measurement errors (yet), but is simply adding in the scatter about the fundamental plane caused by peculiar motions. The exact expression relating this quantity to the true effective radius of the galaxy is
\be
r_{tz}=r_{t}+\eta_t-log_{10}(1+v_{p}/c),
\ee
where $\eta_{t}$ is the true log-distance ratio of the mock galaxy and $v_{p}/c$ is the galaxy's peculiar velocity (normalised by the speed of light) along the observers line-of-sight. This expression follows from the relationship between angular diameter distance, effective radius and comoving distance in Section 4 of \cite{2014MNRAS.445.2677S}.}
\item{Measurement errors on $r_{tz}$ and $i_t$ are generated using the apparent magnitude (derived from $l$ and including the effects of k-correction, surface brightness dimming and galactic extinction) as in \cite{2012MNRAS.427..245M}. The error on the velocity dispersion $\epsilon_{s}$ for each galaxy is generated from a fit of $\epsilon_{s}$ as a function of $s$ itself in the 6dFGSv data. For more details, see Appendix~\ref{AP2}.}
\item{Given the above measurement errors $\epsilon_{r}$, $\epsilon_{s}$ and $\epsilon_{i}$, the observed quantities, $r_{o}$, $s_{o}$ and $i_{o}$, for each mock galaxy are randomly generated from a multivariate Gaussian distribution with mean $\{r_{tz},s_t,i_t\}$ and covariance matrix $E_{n}$ from Eq.~13 of \cite{2012MNRAS.427..245M}. The appropriate selection functions are then applied.}
\item{ The mock galaxies are sub-sampled as a function of RA and declination using the angular completeness mask for the 6dFGS survey \citep{2009MNRAS.399..683J}.
They are also subsampled to match the 6dFGSv redshift distribution in the same way as the 2MTF mocks.}
\item{From the remaining observed mock $r_{o}$, $s_{o}$ and $i_{o}$ the measured log-distance ratio and associated error are generated using the same fitting procedure as was used for the 6dFGSv data in \cite{2014MNRAS.445.2677S}, including the correction for Malmquist bias.}
\end{enumerate}

In Fig.~\ref{logdmock}, we plot the distribution of $\eta$ for five example 6dFGSv mocks alongside the real 6dFGSv data. The distribution of log-distance ratios in the mocks, after correction for Malmquist bias, is well representative of the distribution of the real data. Also, in Appendix~\ref{AP34s} (Figures~\ref{ssD1} and \ref{ssD3}), we plot the redshift distribution and the sky coverage for the 6dFGSv data and example 6dFGSv mocks to show we can represent the inhomogeneous survey geometry of the 6dFGSv data.
 
\begin{figure*}  
  \includegraphics[width=\columnwidth]{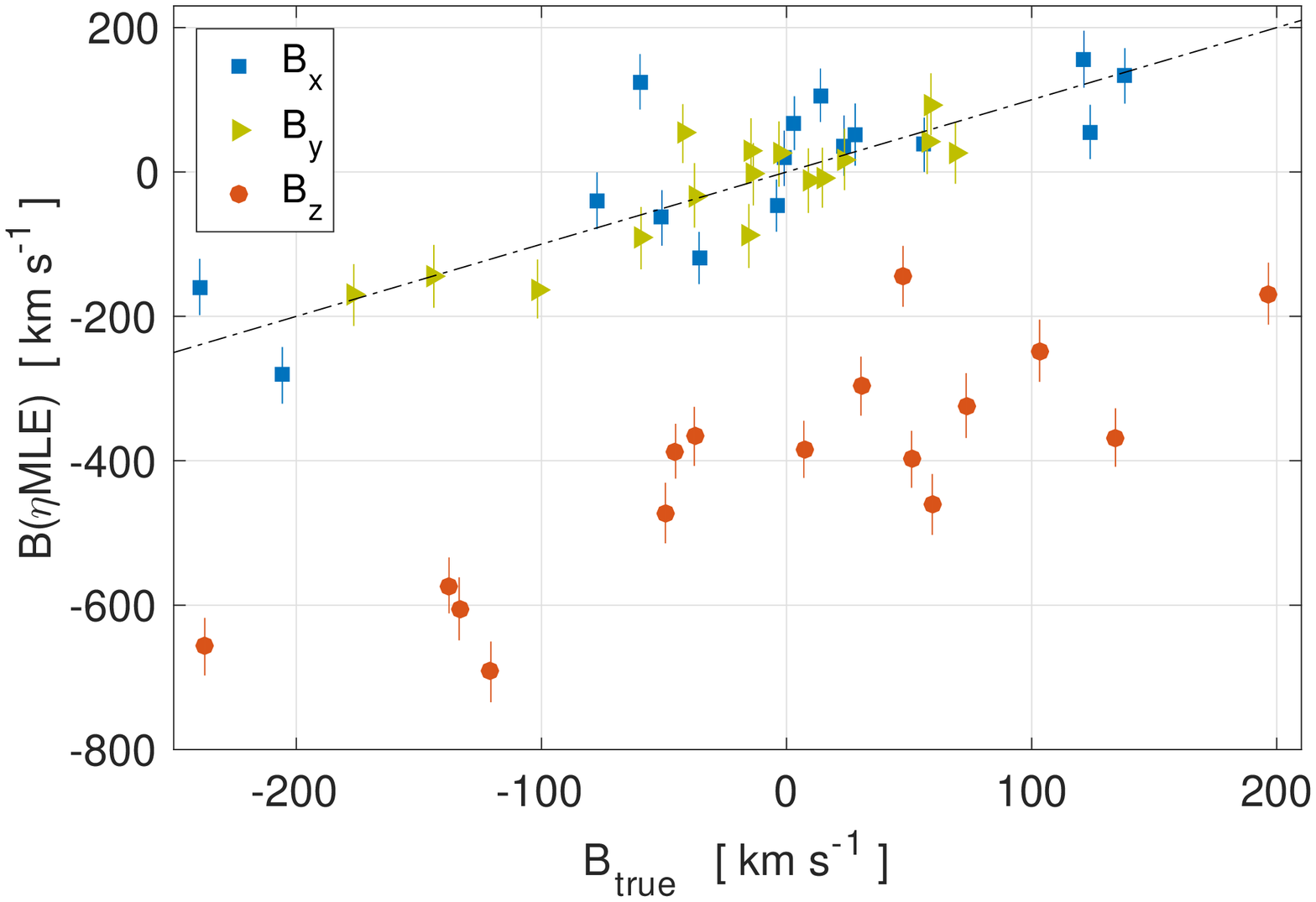}
 \includegraphics[width=\columnwidth]{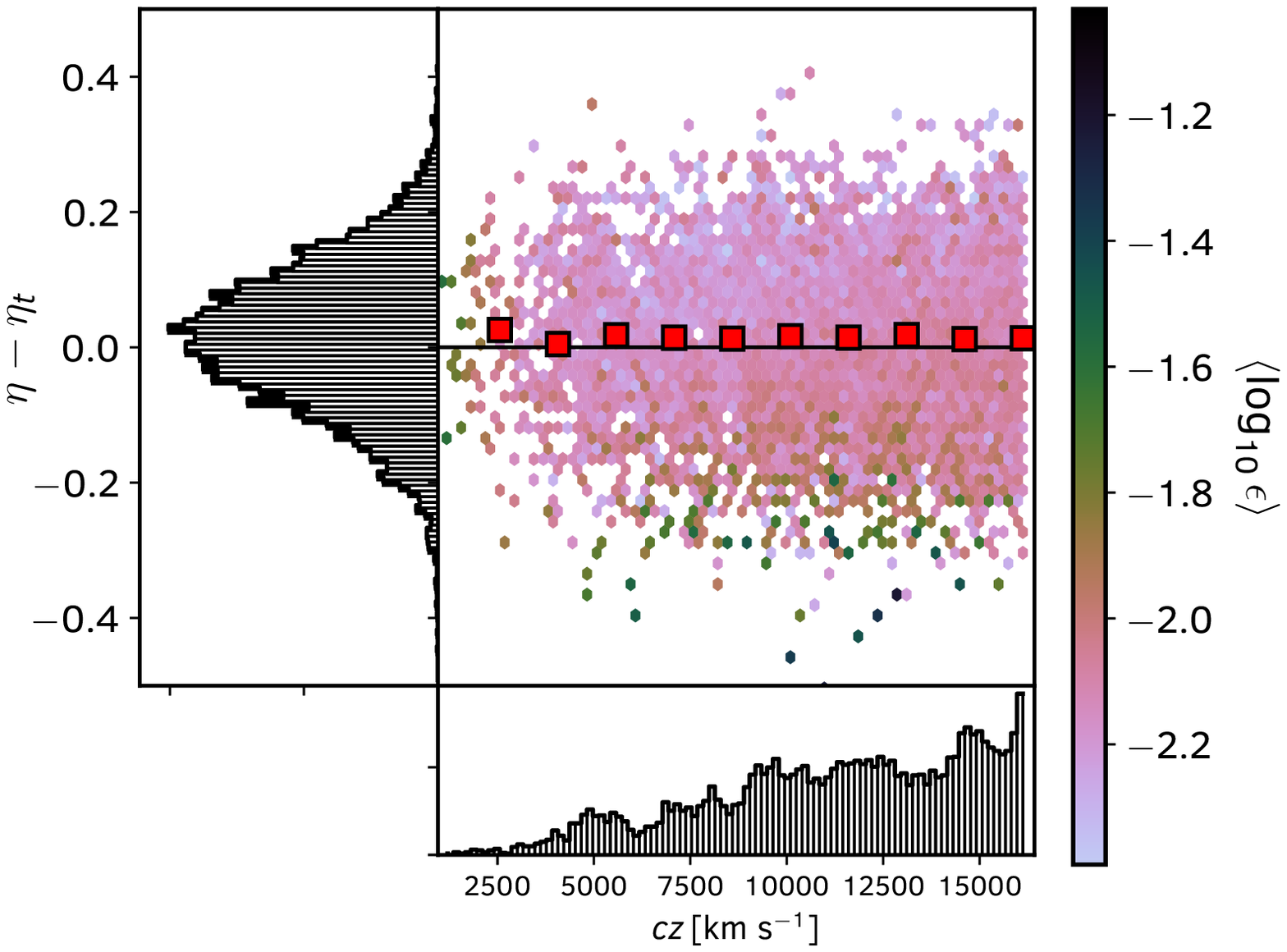}
 \caption{The left-hand panel shows the bulk flow measurement for the original 6dFGSv mocks in equatorial coordinates. In the right-hand panel, we show the difference between the measured logarithmic distance ratio $\eta$ and the true logarithmic distance ratio $\eta_t$ for mock galaxies as a function of redshift. Each point is a `hexbin' colour-coded by the average error $\epsilon^2=\epsilon^{2}_{\eta}+\epsilon^{2}_{\star}$ of the galaxies in that bin to highlight the contribution of different regions of the $\eta-\eta_t$ vs. redshift space to the bulk flow measurement. Side panels are histograms over all galaxies. The red squares are the weighted mean of $\eta-\eta_t$ in redshift bins, weighted by $1/\epsilon^2$. $\eta-\eta_t$ is significantly larger than 0 (compared to the standard error) for nearly all bins.}
\label{6dfmock} 
\end{figure*}

The bulk flow in equatorial coordinates for the 6dFGSv mocks, measured using the $\eta$MLE,  is shown in left-side panel of Fig.~\ref{6dfmock} and compared to the true bulk flow in each mock. We see that the bulk flow in the $x$ and $y$ directions is well recovered by the estimator, but that the $B_z$ component of bulk flow velocities are systematically negative. An important point is that, whilst the true bulk flow in the real 6dFGSv data is unknown, we find a similar amplitude in this direction ($B_z=-439\pm38\,\mathrm{km\,s^{-1}}$), which leads us to believe the result from the real data is also likely to be biased. We find that the origin of this bias arises from a combination of an imperfect correction for Malmquist bias in the data and mocks which is exacerbated by the hemispherical nature of the survey. This bias also occurs when using the $w$MLE and the Minimum Variance estimator, as shown in Appendix~\ref{AP3}.

\begin{figure}  
  \includegraphics[width=\columnwidth]{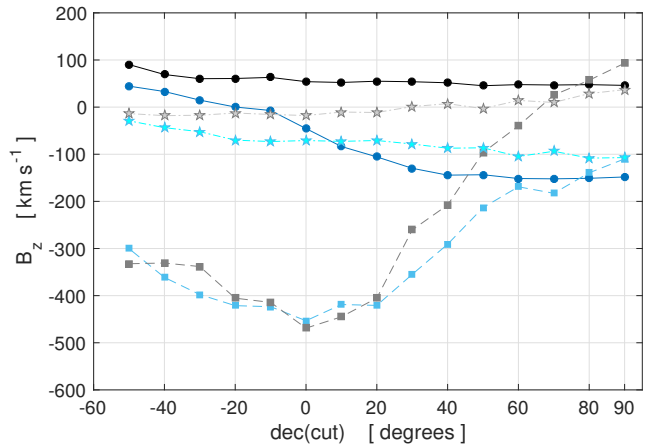}
 \caption{The $z$-component of the bulk flow velocity $B_z$ as a function of declination cut-off dec(cut) for two example 6dFGSv mocks. The filled circles ($\bullet$) represent the true $B_z$ values; stars ($\star$) represent the $B_z$ values estimated using $\eta$MLE and the true log-distance ratio $\eta_t$; and squares ($\square$) represent the $B_z$ values estimated using $\eta$MLE and the `measured' log-distance ratio $\eta$. The blue colors are for one example mock, the black colors are for a second example mock.}
\label{BzDec}
\end{figure}

Firstly, as a demonstration of the fact that this bias is linked to the hemispherical nature of the 6dFGSv survey and the way in which log-distance ratios are measured from the FP, we look at the measured $B_z$ component in two of our mocks when different declination cuts are applied. The mocks are otherwise identical to those used in the rest of this work. We increase the number of galaxies in proportion to the surface area of the survey, such that a hemispherical mock including the zone of avoidance (ZoA) contains the same number of galaxies as 6dFGSv, but the full-sky mock has $\sim 2.4$ times as many (The extra $0.4$ comes from the inclusion of mock galaxies in the ZoA). The results are shown in Fig.~\ref{BzDec}. 

We find that for both the mocks (with different true bulk flows), the result calculated using the true log-distance ratio and the $\eta$MLE method is consistent with the true bulk flow regardless of the declination cut applied. When measured log-distance ratios are used we find that the z-direction bulk flow measurements become increasing biased as we go from a full-sky to hemispherical survey.

To explore this further, we look at the distribution of measured log-distance ratios minus the true log-distance ratio or each galaxy in the mocks. As shown in the right-side panel of Fig.~\ref{6dfmock}, we find that the method used to convert the measured fundamental plane parameters for each mock galaxy to a measured log-distance ratio from \cite{2014MNRAS.445.2677S} is slightly biased, producing on average log-distance ratios that are larger than the true values. This is apparent in the overall histogram of differences, and when looking at the weighted average in redshift bins (where each galaxy is weighted by $1/(\epsilon_{\eta}^2+\epsilon_{\star}^2)$). Upon further investigation, we find that this is caused by the normalisation of the log-distance probability
distribution function (PDF) for each galaxy, namely `$f_n$' in equation 5 of \cite{2014MNRAS.445.2677S} which also attempts to correct for Malmquist bias. 

As stated in \cite{2014MNRAS.445.2677S} this normalisation is computed numerically using Monte Carlo samples of fundamental plane parameters drawn from the best-fit 6dFGSv fundamental plane and applying the magnitude limit of the 6dFGSv sample ($J<13.65$ mag). By minimizing the $\chi^2$ difference between the true and measured $B_z$ for each of our 16 mocks as a function of this magnitude limit we find we are able to remove the bias in the z-direction bulk flow measurement of 6dFGSv if a best-fit value of $J<13.217$ mag is used in the `$f_{n}$' calculation instead. Why the best-fit value differs from the magnitude limit expected for the 6dFGSv data is unclear and would involve a detailed look at the 6dFGSv photometry, photometric errors and completeness, which is beyond the scope of this work.

\begin{figure*}  
  \includegraphics[width=\columnwidth]{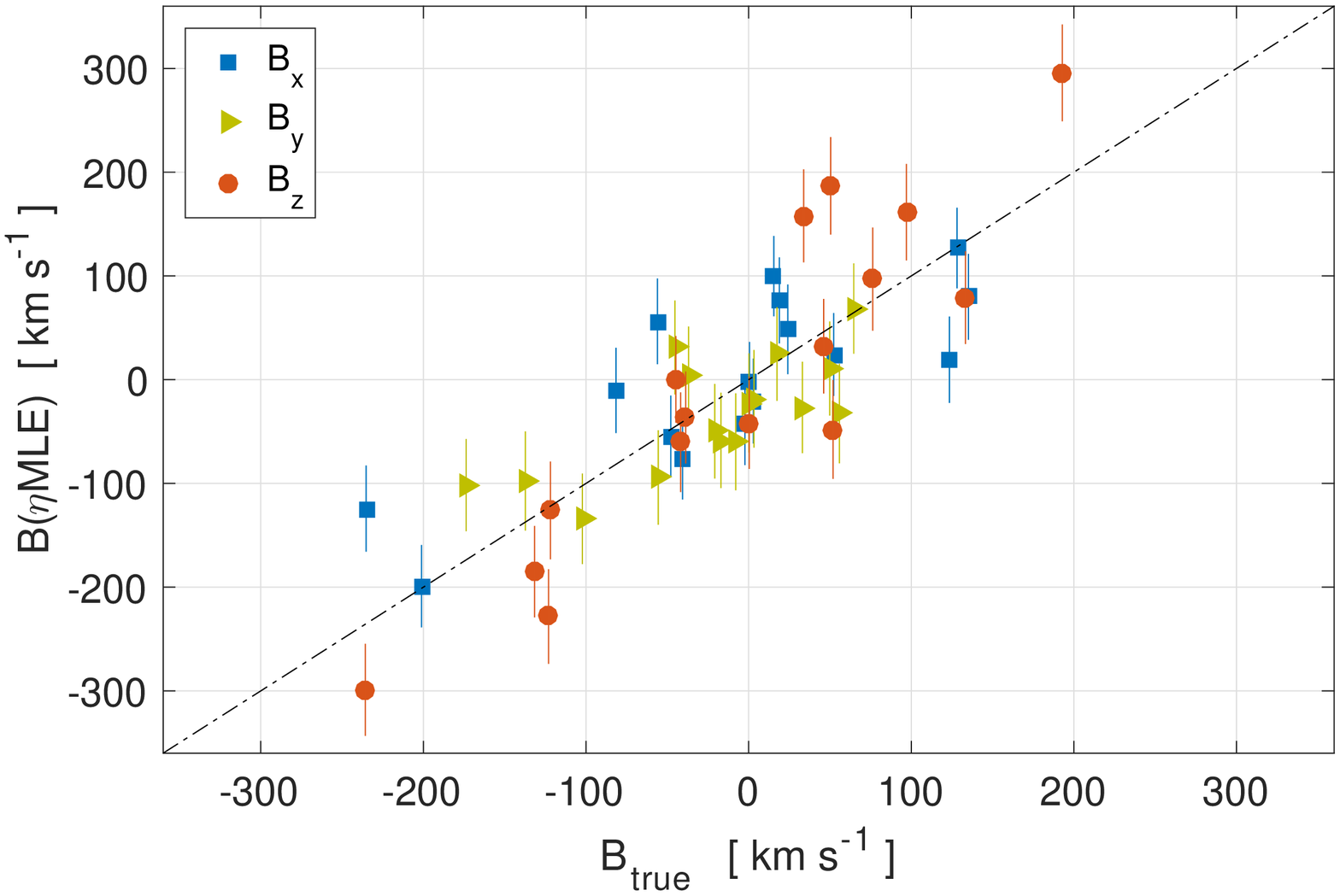}
 \includegraphics[width=\columnwidth]{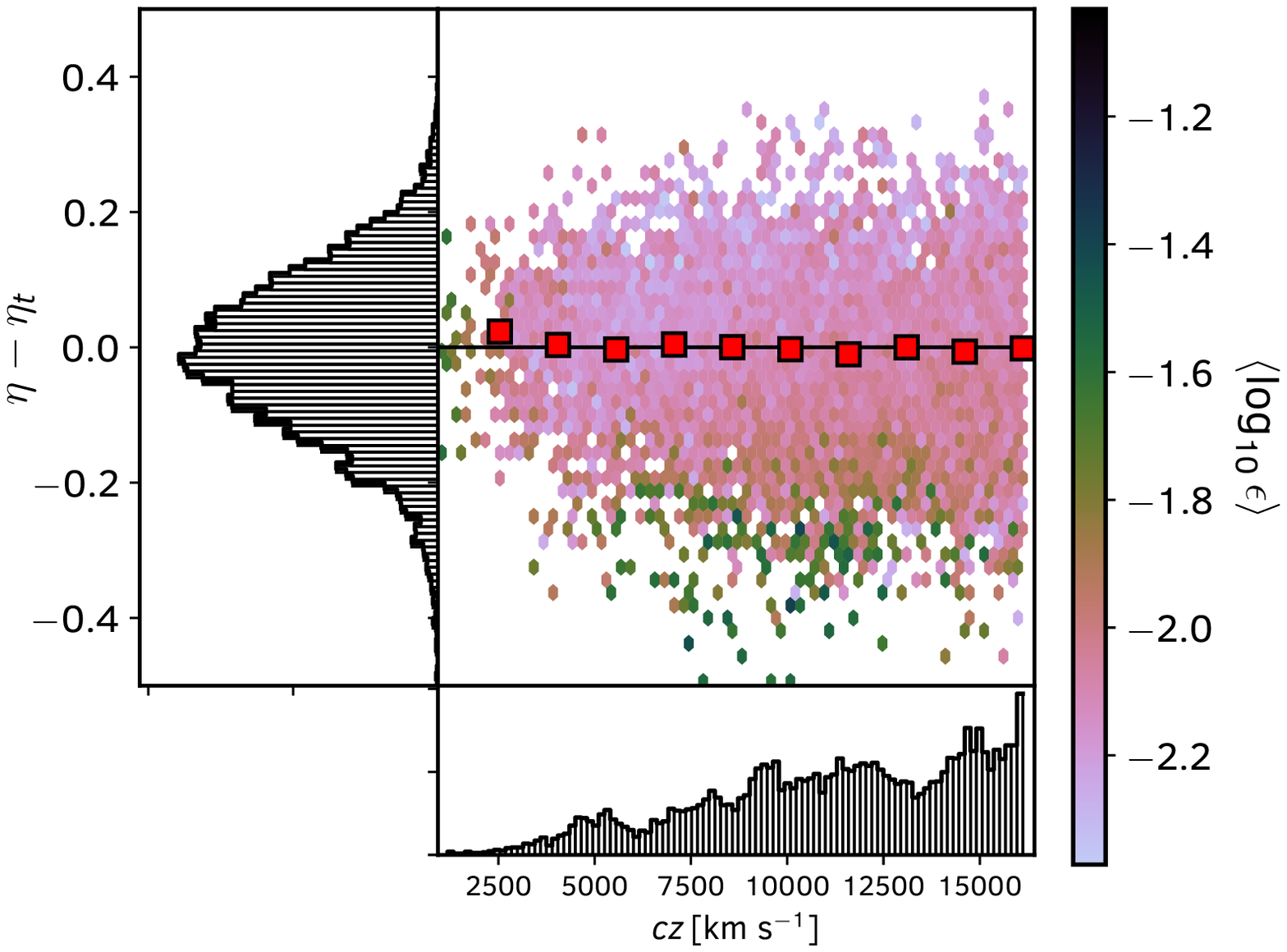}
 \caption{Same as Fig.~\ref{6dfmock} but using the re-calibrated `$f_{n}$' for each galaxy. We find the bias in $B_{z}$ is now effectively removed, and $\eta-\eta_{t}$ is consistent with zero across all redshift bins.}
\label{6dfmock2}
\end{figure*}

However, using the re-calibrated `$f_n$' to calculate the log-distance ratio, and then the bulk flow of the 6dFGSv mocks, we recover the results shown in left-side panel of Fig.~\ref{6dfmock2}. In the right-side panel of Fig.\ref{6dfmock2}, we demonstrate that this correction has effectively removed the difference between the true and measured log-distance ratios in the mocks. Finally, we also use this re-calibrated $f_n$ to calculate the bulk flow of the 6dFGSv data, and the results are shown in Table~\ref{bkflradec}.

An alternative way to correct for the $B_{z}$ bias without exploring the data in detail would be to calculate the difference in the true and measured $B_{z}$ averaged over the mocks and apply this directly to the computed 6dFGSv $B_{z}$ value. This method is used in Appendix~\ref{AP3} and gives nearly identical results to the `$f_{n}$' correction. This gives us confidence that, although the reasons for re-calibration of `$f_{n}$' are not fully understood, the correction itself is robust and accurate.

\subsection{Combined mocks}

\begin{figure}  
 \includegraphics[width=\columnwidth]{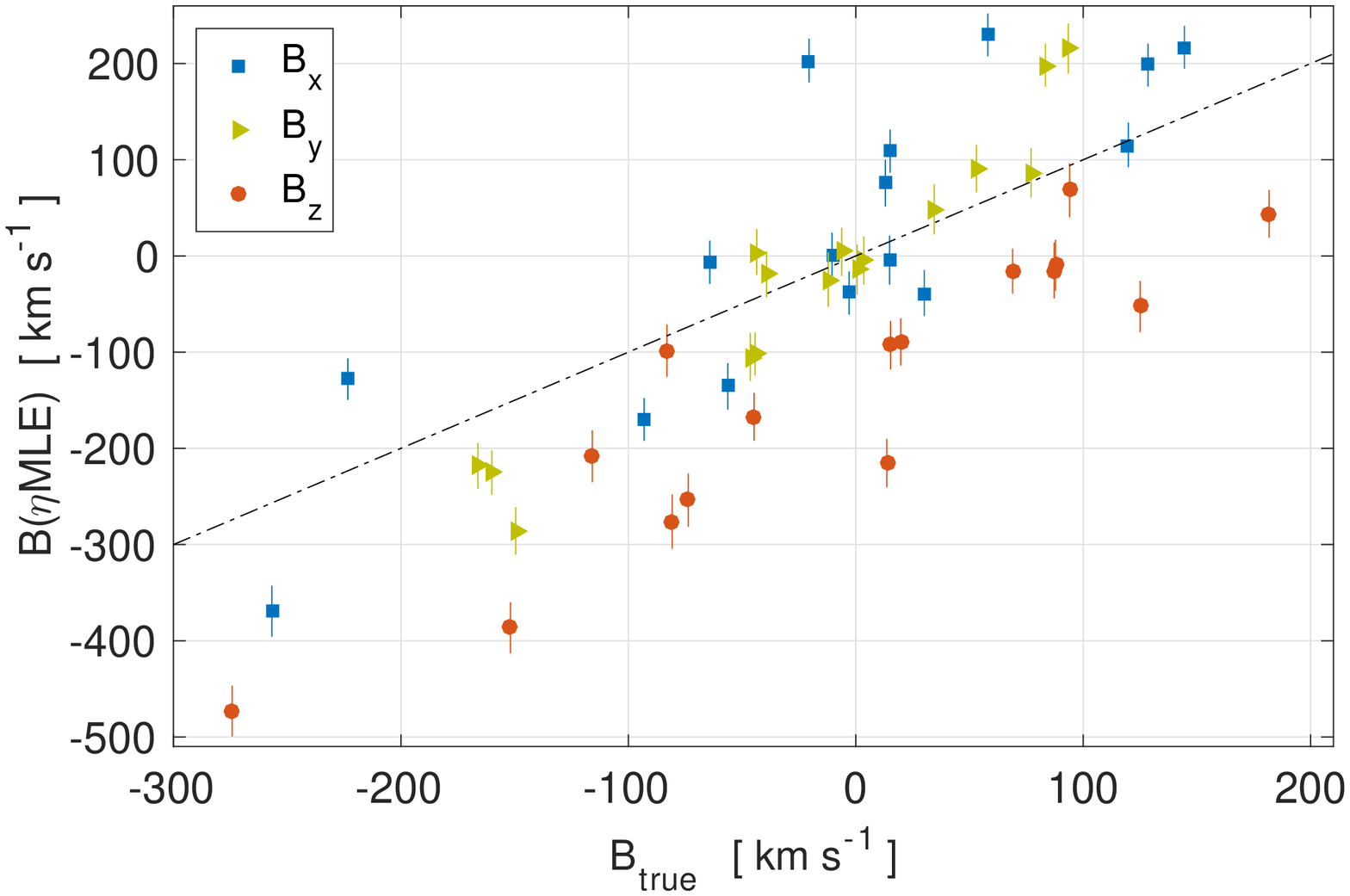}\\
 \includegraphics[width=\columnwidth]{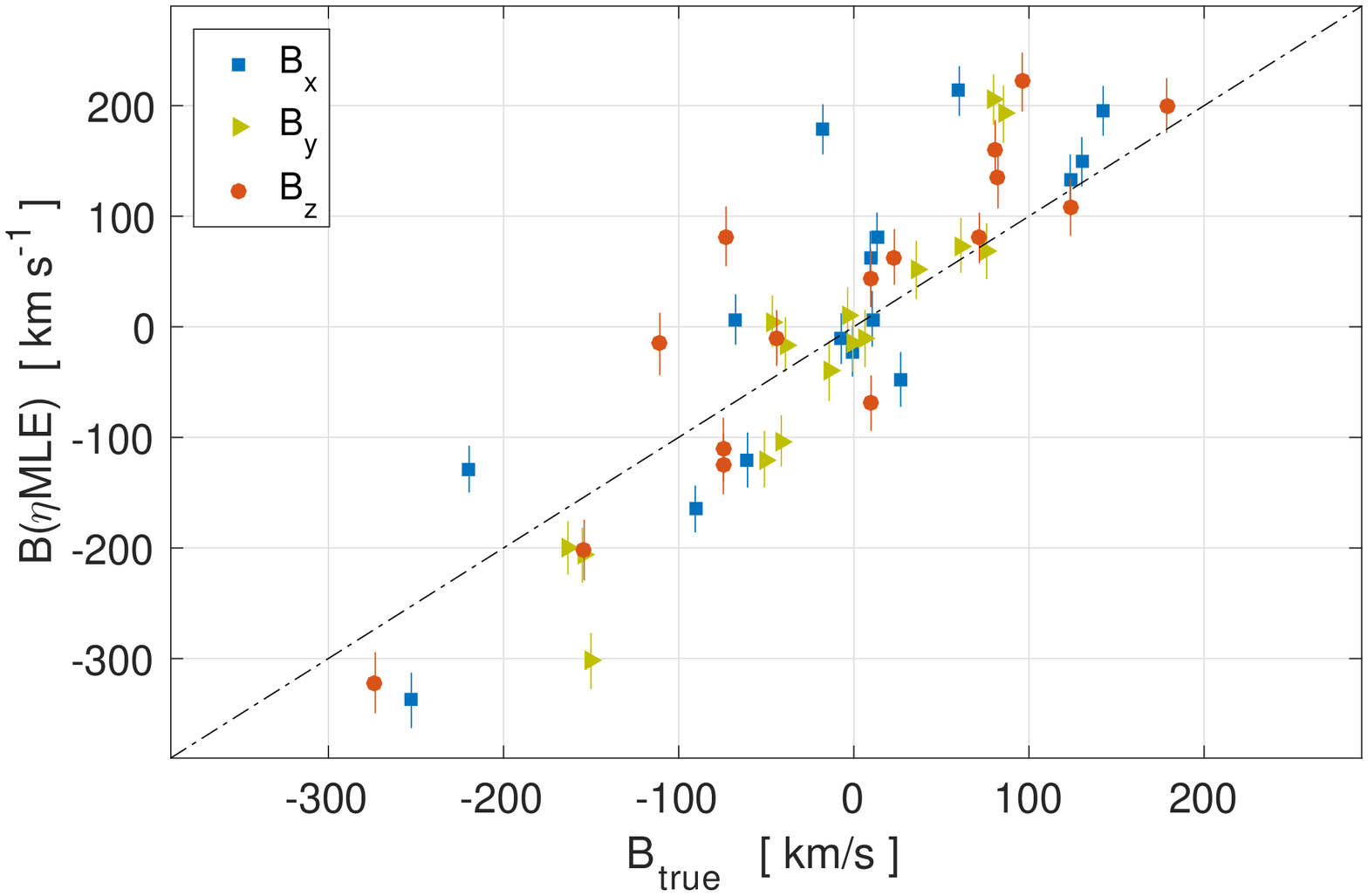}
 \caption{The bulk flow measurement for the combined mocks in equatorial coordinates. The black dashed line represents equality. The top panel shows the (biased) results using the original 6dFGSv mocks, whilst the bottom panel shows the bias-free `$f_n$'-revised bulk flow.}
\label{CBmock}
\end{figure}

As both the 2MTF and 6dFGSv mocks reproduce the respective selection functions of the two dataset and are centred on the same observer, they can be combined in the same manner as the data to produce a set of $2 \times 8$ combined mock galaxy catalogues. We use these to test the expected performance of the $\eta$MLE fitting method on the combined sample. The measured bulk flow components for the combined mocks are shown in Fig.~\ref{CBmock}.
 
Compared to the (biased) 6dFGSv-only mocks, the bias in the z-direction bulk flow velocities $B_{z}$ is reduced significantly due to the more isotropic distribution of galaxies. However, some bias remains as the number density of the galaxies in southern sky is still much higher than in the north sky (by a ratio of $\sim 5:1$). If we combine the `$f_n$'-revised 6dFGSv mocks with the 2MTF mocks, the bias of $B_z$ again vanishes.

\section{Results and discussion}

\subsection{Bulk flow results}\label{reslts}

The results for our fits to the bulk flow in equatorial coordinates for the combined and individual 2MTF and 6dFGSv datasets are presented in Table~\ref{bkflradec}, while the results in Galactic coordinates are presented in Table~\ref{bkflb}.

The error in the 2MTF bulk flow is smaller than 6dFGSv 
even though the 6dFGSv data has more galaxies. This is due to the smaller distance error of the 2MTF data (both in the fractional sense, and because the galaxies are closer). The error of the bulk flow in the combined data set improves over this further, giving a reduction of around $\sim 30\%$, similar to the forecast improvement in the growth rate constraints from a combined dataset \citep{2017MNRAS.464.2517H}. This is not suprising given that, in the absence of systematic modelling errors, the same properties (number density and typical distance error) determine the accuracy with which both the growth rate and bulk flow can be measured.

For the measurement error of the 6dFGSv bulk flow amplitude, the $\eta$MLE result is around $\pm 35$ km s$^{-1}$, while in Table 1 of \cite{2016MNRAS.455..386S}, the MV and the $d$MLE error is around $\pm 50$ km s$^{-1}$, which is larger than our $\eta$MLE result. This is not due to the $\eta$MLE method, (our $w$MLE result is $240.6\pm36.3$ $km/s$, which is similar to the $\eta$MLE results), but rather the more accurate way in which both of these newer methods account for the measurement error on each galaxy's velocity.

Both the $\eta$MLE and the $w$MLE can convert the measurement errors of the log-distance ratio, $\epsilon_{\eta}$ to the measurement errors of the bulk flow in a way than encapsulates the non-Gaussian nature of the errors in velocity space. The $w$MLE simply allows us to convert $\epsilon_{\eta}$ to the measurement errors of peculiar velocities, $\sigma_{v}$ through an analytic relation inferred from Eq.\ref{watvp}, then to the measurement error of bulk flow. In $\eta$MLE, we can calculate the measurement errors of the bulk flow velocities directly from $\epsilon_{\eta}$ through the MCMC chains generated under the likelihood function of $\eta$ (i.e.  Eq.~\ref{pvpi}) rather than via $\sigma_v$.

By contrast, \cite{2016MNRAS.455..386S} use a fit (Fig. 4 therein, or our Eq.~\ref{appdcz1}) for both their MV estimator and $d$MLE to calculate the measurement errors of the peculiar velocities. This method does not fully include all aspects of the uncertainty, and gives larger errors on each galaxy. As a result, in the final bulk flow measurement, both the MV and the $d$MLE have larger measurement errors.

\begin{table*}   \small
\caption{Bulk flow measurements in equatorial coordinates. }
\begin{tabular}{|c|c|c|c|c|c|c|c|}
\hline
\hline
\multicolumn{8}{|c|}{Equatorial Coordinates}\\
\hline

Data set   & $|\vec{B}|$ & $B_x$ &   $B_y$  &  $B_z$ 
& RA & Dec & Depth \\

&km s$^{-1}$ &   km s$^{-1}$  & km s$^{-1}$&km s$^{-1}$
& degree&degree &Mpc h$^{-1}$\\
\hline

2MTF& $368.8\pm32.5$  &$-228.3\pm26.4$    & $21.6\pm35.5$ &  $-285.8\pm33.1$ &    $174.6\pm 8.8$&  $-51.3\pm 4.3$   & 29.1 \\


6dFGSv& $233.3\pm35.4$  &$-214.8\pm35.9$    & $-83.6\pm45.1$&  $-36.1\pm43.0$ &    $201.3\pm 11.2$&$-8.9 \pm 10.4$     & 75.6 \\


Combined& $287.6\pm23.8$  &$-214.4\pm21.3$    & $-16.9\pm28.0$ &  $-190.9\pm25.3$ &    $184.5\pm 7.5$&$-41.6\pm  4.5  $ & 38.9 \\
\hline
          
\end{tabular}
 \label{bkflradec}
\end{table*}

\begin{table*}   \small
\caption{Bulk flow measurements in Galactic coordinates.  
}
\begin{tabular}{|c|c|c|c|c|c|c|c|}
\hline
\hline
\multicolumn{8}{|c|}{Galactic Coordinates}\\
\hline
 
Data set   & $|\vec{B}|$ & $B_x$ &   $B_y$  &  $B_z$ 
& $\ell$ & $b$ & Depth \\

&km s$^{-1}$ &   km s$^{-1}$  & km s$^{-1}$&km s$^{-1}$
& degree&degree &Mpc h$^{-1}$\\
\hline

2MTF&   $369.2\pm32.5$    & $131.9\pm36.4$ &   $-335.9\pm33.2$ &  $63.5\pm25.5$ &$291.5\pm 6.2$&   $9.9\pm 4.1$  &29.1\\

6dFGSv& $233.3\pm35.1$  &$102.2\pm43.2$    & $-95.8\pm42.3$ &  $186.5\pm38.2$ &    $316.8\pm 21.6$& $53.1\pm 10.1$     & 75.6 \\

Combined& $287.6\pm23.9$  &$118.9\pm28.0$    & $-241.0\pm25.6$ &  $102.3\pm21.2$ &    $296.3\pm 5.7$& $20.8\pm 4.6$    & 38.9 \\
\hline
          
\end{tabular}
 \label{bkflb}
\end{table*}

\subsection{Comparison with theory and previous results} \label{sec:theorycomp}

In this section we compare our measurements from the combined sample to the predictions from linear theory and results from other datasets.

At redshift zero, under the $\Lambda$CDM model and assuming General Relativity, the growth factor $f=\Omega_m^{0.55}$, and the variance of the bulk flow velocity is \citep{2012ApJ...761..151L,2014MNRAS.445..402H,2016MNRAS.463.4083A}:
\be\label{sigB}
\sigma_B^2=\frac{H_0^2f^2}{2\pi^2}\int\mathcal{W}^2(k)\mathcal{P}(k)dk
\ee
where $\mathcal{P}(k)$ is the linear matter density power spectrum (which we generate using the CAMB package \citealt{Lewis:1999bs, 2012JCAP...04..027H}) and $\mathcal{W}(k)$ is the Fourier transform of the survey window function. 

\begin{figure} 
\centering
 \includegraphics[width=\columnwidth]{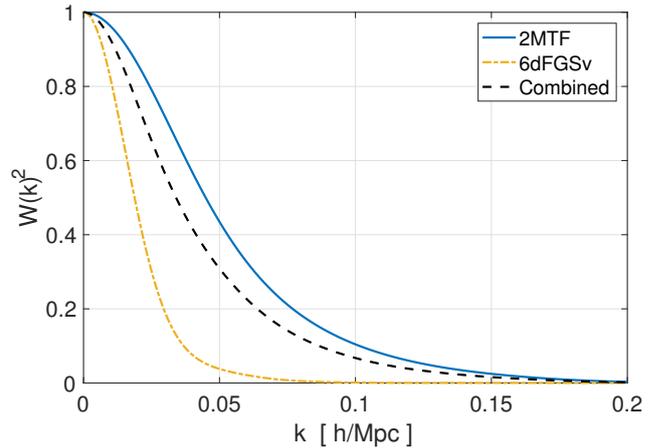}
 \caption{Window functions for 2MTF, 6dFGSv and the combined data set.}
 \label{windCB}
\end{figure}

Commonly used forms of the window function are the Gaussian $\mathcal{W}(k)=\mathrm{exp}(-k^2R^2/2)$ and the spherical top-hat
$\mathcal{W}(k)=3(\sin kR-kR\cos kR)/(kR)^3$ \citep{2012ApJ...761..151L, 2016MNRAS.463.4083A}. However, 6dFGSv is a hemispherical survey rather than full-sky, 2MTF is inhomogenous above and below a declination $\delta=-40.0^{\circ}$, and both surveys do not cover the Galactic plane. Additionally, the window function also depends on the distance error and therefore the weight assigned to each galaxy. Therefore the correct window function is more complicated than a sphere or Gaussian whose outer radius is equal to the bulk flow depth. Instead we generate more accurate window functions that account for the above using the following algorithm:
\begin{enumerate}
\item{Generate $N\times2062$ random points with the same sky and redshift distribution as the 2MTF survey and $N\times8885$ random points with the same sky and redshift distribution as the 6dFGSv survey. We use the same procedure as for the mock catalogues, but do not include other selection effects such as magnitude limits.}
\item{Add these two sets of random points together and convert the ($cz$, RA, Dec) to $(x, y, z)$ coordinates. Producing the two sets separately ensures the same density of points as a function of sky position and redshift as the combined data set.}
\item{Perform the following summation over the random points for a given $k$:
\be\label{windf}
\mathcal{W}(k)=\frac{\sum_{l=1}^{N_{tot}}w_le^{ i\frac{k}{\sqrt{3}}(x_l+y_l+z_l) }}{\sum_{l=1}^{N_{tot}}w_l}
\ee
where $N_{tot}=N\times(2062+8885)$ is the total number of random points and $w_{l}$ are the weights assigned to each random point. We mimic the contribution of each galaxy at a given distance to the bulk flow measurement using $w_l=1/(\sigma_l^2(r)+\sigma_{\star}^2)$ with $\sigma_l$ calculated by using the distance to the random point, and $\sigma_l$ calculated from Eq.~\ref{dcz1} and Eq.~\ref{dcz2} for 6dFGSv and 2MTF, respectively.}
\end{enumerate}
We find that $N=50$ is suitable to obtain convergence in our estimates of the window function down to large $k$ where the window function vanishes. Window functions for the separate 2MTF and 6dFGSv samples are obtained by summing over only the corresponding random points. The three window functions for our data are shown in Fig.~\ref{windCB}, where as expected the window function for the 2MTF has support for larger $k$ (smaller scales) than for 6dFGSv, with the latter survey covering a larger cosmological volume.

The bulk flow amplitude $|\vec{\boldsymbol{B}}|$ is assumed to follow a Maxwell-Boltzmann distribution \citep{2012ApJ...761..151L, 2016MNRAS.463.4083A} and so the most likely bulk flow amplitude is given by $B_p=\sqrt{2/3}\sigma_B$, and the (cosmic) variance of the bulk flow amplitude 
is $B_p$$^{+0.419\sigma_B}_{-0.356\sigma_B}$ ($1\sigma$) and $B_p$$^{+0.891\sigma_B}_{-0.619\sigma_B}$ ($2\sigma$) \citep{2016MNRAS.455..386S}. 
Using Eq.~\ref{sigB} and the $\mathcal{W}^2(k)$ in Fig.~\ref{windCB}, we can calculate the theoretical bulk flow prediction given in Table \ref{bkvst}. We find that all of our bulk flow measurements (for the 2MTF, 6dFGSv and combined samples) are consistent with the predictions from $\Lambda$CDM. 

\begin{table}   \centering
\caption{Comparing the $\eta$MLE measured bulk flow with the $\Lambda$CDM predicted bulk flow. Errors on the $\Lambda$CDM prediction denote the cosmic variance.}
\begin{tabular}{|c|c|c|}
\hline
\hline
Data set &$\eta$MLE&$\Lambda $CDM\\
  &km s$^{-1}$&km s$^{-1}$\\
\hline

2MTF      & $369\pm33$ & 315$^{+161}_{-137}$\\
\\
6dFGSv    & $233\pm35$ & 217$^{+112}_{-95}$ \\
\\
Combined  & $288\pm24$ & 289$^{+148}_{-126}$\\

\hline
\end{tabular}
 \label{bkvst}
\end{table}

\begin{figure*} 
\includegraphics[width=170mm]{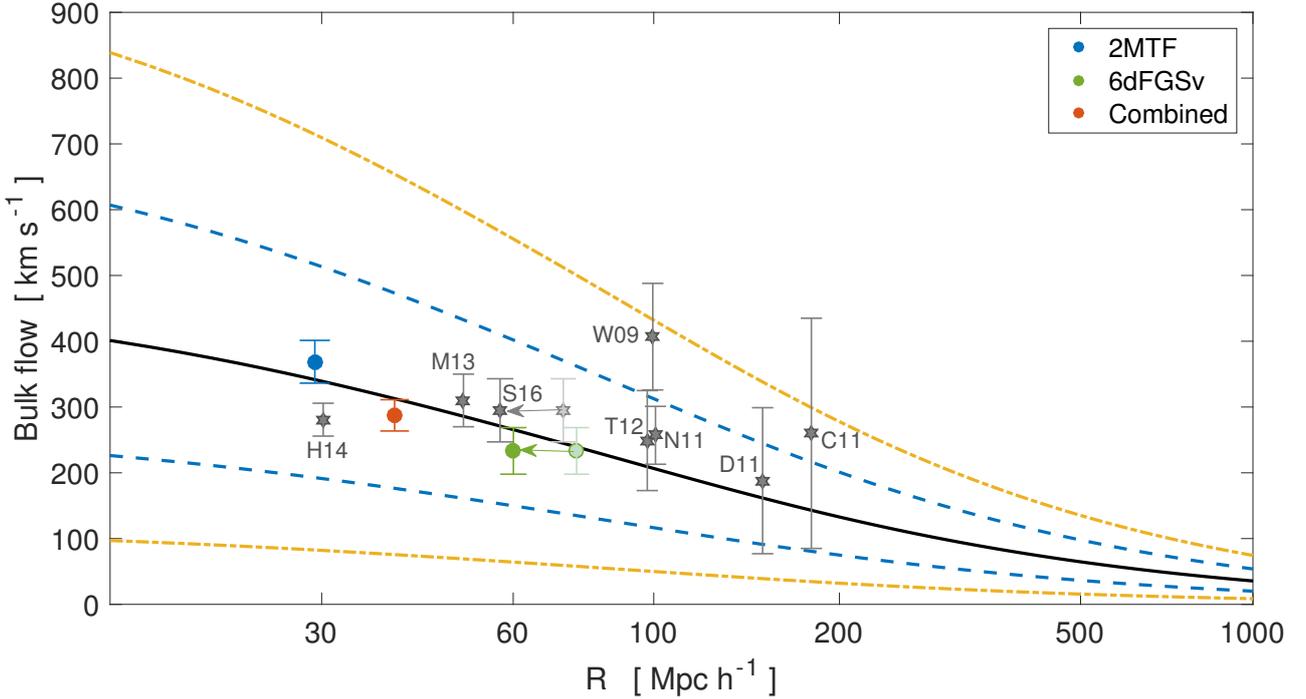}
\caption{Comparison of bulk flow amplitude to the prediction of $\Lambda$CDM. The theoretical model assumes a top-hat window function. The solid line indicates the most probable bulk flow and the blue (yellow) dashed lines indicate the $1\sigma$ ($2\sigma$) values. Filled circles ($\bullet$) are our $\eta$MLE estimated bulk flows (Table \ref{bkflradec}) for 2MTF, 6dFGSv, and the combined data set.  Other recent measurement are shown as gray stars ($\star$) (H14: \citet{2014MNRAS.445..402H}; W09: \citet{2009MNRAS.392..743W}; S16: \citet{2016MNRAS.455..386S}; C11: \citet{2011MNRAS.414..264C}; T12: \citet{2012MNRAS.420..447T}; N11: \citet{2011ApJ...736...93N}; D11: \citet{2011JCAP...04..015D}; M13:
\citet{2013MNRAS.428.2017M}). W09 and T12 use Gaussian windows, and so we plot them at twice their quoted radius, to be comparable to the top-hat window prediction. S16 uses effective radii -- the gray arrow shows how far we have shifted the point from the measured radii. Similarly the green arrow for the $\eta$MLE-measured 6dFGSv data point.}
\label{BvsR} 
\end{figure*}

Because of the differing geometries and depths, it is difficult to compare bulk flow measurements between different surveys. Instead, to visualise how the expected and measurement bulk flow changes over distance, we compare our results and those from other surveys \citep{2014MNRAS.445..402H,2009MNRAS.392..743W,2016MNRAS.455..386S,2011MNRAS.414..264C,2012MNRAS.420..447T,2011ApJ...736...93N,2011JCAP...04..015D,2013MNRAS.428.2017M} to the $\Lambda$CDM predictions for spherical top-hat window functions of different radii.
We emphasise that the correct comparison between the measurement from a particular dataset and the theoretical prediction of $\Lambda$CDM should properly account for the (possibly complicated) survey window function, as in Table~\ref{bkvst}. However, for the purposes of comparison with other datasets, it is necessary to standardise this window function. We have chosen the spherical top-hat window for its simplicity and, where necessary, converted the results of studies with different window functions following the argument in \cite{2016MNRAS.455..386S}, placing these measurements at distances that differ from the actual bulk flow depth given by the original authors. The \cite{2009MNRAS.392..743W} and \cite{2012MNRAS.420..447T} measurements have Gaussian windows, and so in Fig.~\ref{BvsR}, we plot them at twice their quoted radius to be more comparable to the spherical top-hat window prediction. The 6dFGSv data is a hemispherical top-hat so we plot both the results from \cite{2016MNRAS.455..386S} and this work at the bulk flow depth, with an arrow to indicate how these results move if we use a smaller effective radius $R_{eff}=(R^3/2)^{1/3}$. All other surveys, including the 2MTF and combined sample here have a window function close to a spherical top-hat and are placed at their stated bulk flow depth.
From Fig.~\ref{BvsR}, we can see 
the majority of measured bulk flows are in agreement with the predictions from $\Lambda$CDM at the $1\sigma$ level, while W09 and C11 is in agreement with $\Lambda$CDM at the $2\sigma$ level. The $\eta$MLE-measured bulk flows have significantly smaller errors than the others. Our combined measurement is the most accurate bulk flow measurement to date.

\begin{figure*} 
\includegraphics[width=175mm]{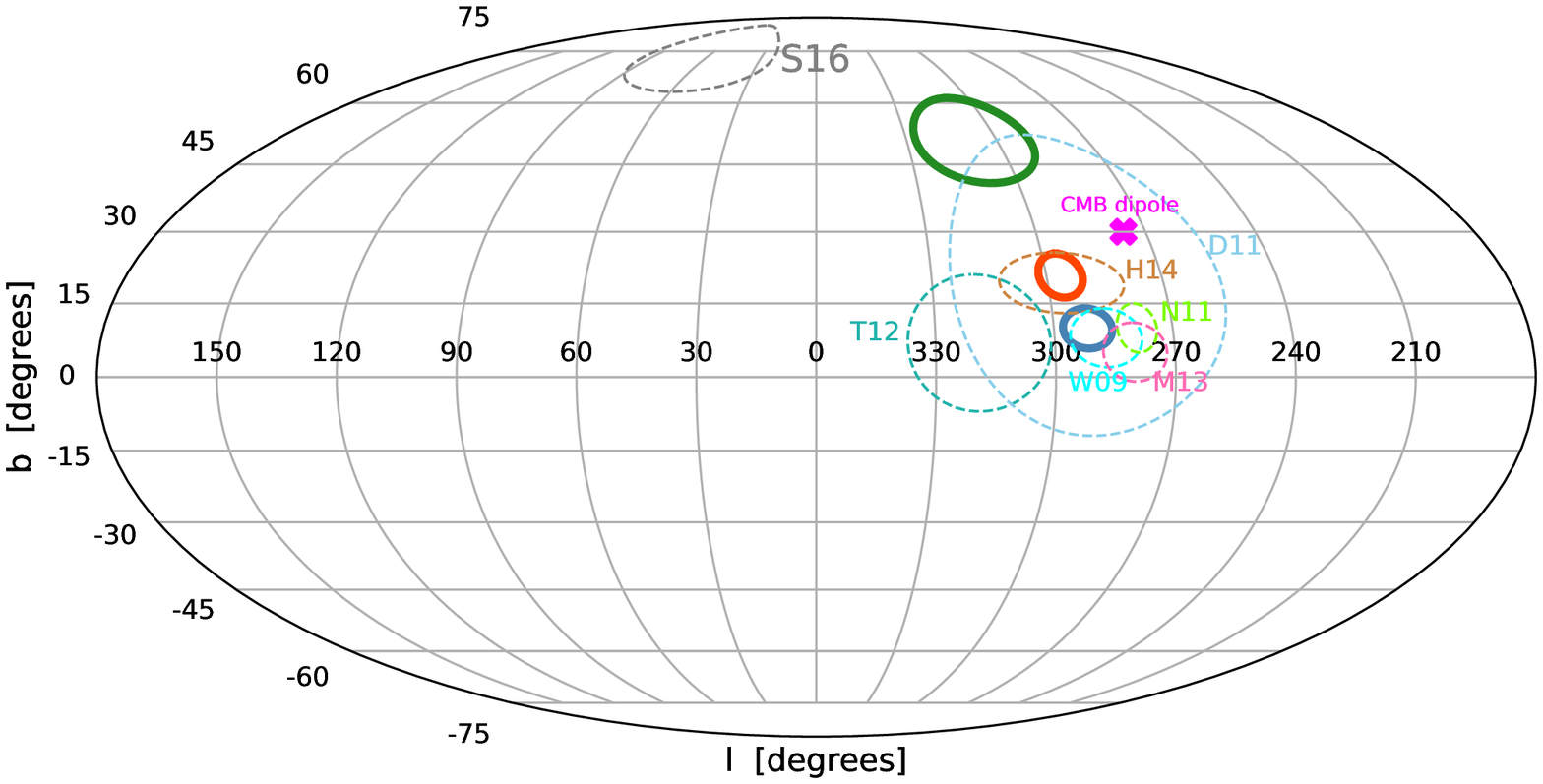}
\caption{Comparison of the bulk flow direction in Galactic coordinates. The blue, green and the red solid circles are the $\eta$MLE measured bulk flows for 2MTF, 6dFGSv and the combined data sets, respectively.  Other recent measurements are shown as the colored dashed circles: (H14: \citet{2014MNRAS.445..402H}; W09: \citet{2009MNRAS.392..743W}; S16: \citet{2016MNRAS.455..386S}; T12: \citet{2012MNRAS.420..447T}; N11: \citet{2011ApJ...736...93N}; D11: \citet{2011JCAP...04..015D}; M13:
\citet{2013MNRAS.428.2017M}. The size of the circles indicates the $1\sigma$ error. The pink cross is the direction of the CMB dipole.}
\label{lbcomp} 
\end{figure*}

In Fig.~\ref{lbcomp}, we compare the bulk flow direction in Galactic coordinates. 
The bulk flow directions are mainly in agreement with other authors' results. However, the bulk flow direction of S16 (measured using the $d$MLE) is different from other results. Our new 6dFGSv-only measurement is also in disagreement, but less so. This is likely due to cosmic variance arising from the depth and sky coverage of 6dFGSv and, in the case of S16, the effect of selection and hemispheric bias that was not accounted for. The bulk flow direction seem to be converging towards the CMB dipole \citep{2011MNRAS.414..264C}, but it is possible that the amplitude may not dip below 150 km~s$^{-1}$ until depths of 200-500 Mpc h$^{-1}$ are reached. This is beyond the distance of the Shapley supercluster, which is undoubtedly responsible for some of the bulk flow.

\section{Conclusions}

We have used the 2MTF and 6dFGSv surveys, individually and combined, to the measure bulk flow of galaxies within the local Universe. The combined sample in particular increases the effective depth and volume of 2MTF alone, and vastly improves on the hemispherical bias of 6dFGSv alone. 

We demonstrate the extent that this bias has afflicted previous measurements. We also investigate the effect of the mainly log-linear measurement errors on the accuracy of different Maximum Likelihood Estimates (MLEs), including a `magnitude fluctuation' estimator $\eta$MLE which calculates the full probability density function for each bulk flow possibility, and can thus deal with a mixture of linear error terms (e.g. non-linear velocity dispersion) and log-linear terms (e.g. distance errors).

We test the MLE methods and explore the effect of sample selection by constructing 16 independent mock surveys from the large-scale GiggleZ and SURFS simulations. A high degree of consistency is shown between different estimators, with the $\eta$MLE technique being more accurate, especially compared with MLE methods which assume Gaussian errors in peculiar velocity. 

The uncertainty in the equatorial $z$-component of bulk flow for 6dFGSv is shown to be greater than previous studies have suggested. Using mock surveys, we have identified and corrected for a bias in the 6dFGSv measurements caused by systematic errors in the Malmquist bias correction in the data and related to the photometric properties of the sample. We have explored the magnitude of this bias as a function of sky coverage in order to inform the design of future sky surveys from single ground-based sites such as the Taipan Galaxy Survey \citep{2017PASA...34...47D}, SkyMapper \citep{2018arXiv180107834W} and LSST \citep{2008arXiv0805.2366I}. Correcting for Malmquist bias in peculiar velocity surveys is difficult and there is often the potential for unknown systematics. Our investigation here has shown that hemispherical surveys are particularly susceptible, and so greater care must be taken in analysing future surveys such as the Taipan Galaxy Survey than for more isotropic surveys such as WALLABY \citep{2012PASA...29..359K}.

Using the individual and combined 2MTF and 6dFGSv samples, we measure bulk flow amplitudes (depths) of $369\pm33$ km s$^{-1}$ ($29h^{-1}$ Mpc), $233\pm35$ km s$^{-1}$ ($76h^{-1}$ Mpc), and $288\pm24$ km s$^{-1}$ ($39h^{-1}$ Mpc), respectively. All values are consistent with the $\Lambda$CDM expectation values of 315$^{+161}_{-137}$ km s$^{-1}$, 217$^{+112}_{-95}$ km s$^{-1}$, and 289$^{+148}_{-126}$ km s$^{-1}$, respectively.

\section*{Acknowledgements}

This research was conducted by the Australian Research Council Centre of Excellence for All-sky Astrophysics (CAASTRO), through project number CE110001020.

We thank Greg Poole for providing access to the GiggleZ simulation and Pascal Elahi, Chris Power, Claudia Lagos, Aaron Robotham and Rodrigo Ca\~nas for producing the SURFS simulation and associated products. The SURFS simulation suite was undertaken on Magnus at the Pawsey Supercomputing Centre in Perth, Australia and on Raijin, the NCI National Facility in Canberra, Australia, which is supported by the Australian commonwealth Government.

This work used data from the Robert C. Byrd Green Bank Radio Telescope obtained through observing projects GBT06A-027, GBT06B-021, GBT06C-049, GBT08B-003: ``Mapping Mass in the Nearby Universe with 2MASS'', PI Karen L. Masters.

This research has made use of NASA's Astrophysics Data System Bibliographic Services, the \texttt{astro-ph} pre-print archive at \url{https://arxiv.org/} and the {\sc matplotlib} plotting library \citep{2007CSE.....9...90H}.




\bibliographystyle{mnras}
\bibliography{BKFmn} 



\appendix

\section{2MTF PECULIAR VELOCITY measurement errors for \lowercase{d}MLE}\label{AP1}

Due to the non-linear nature of the transformation from the observed log-distance ratio $\eta$, to a peculiar velocity $v$, there is some choice to be made about how to quantify the errors in the peculiar velocity measured for each galaxy. 

\begin{figure}  
 \includegraphics[width=\columnwidth]{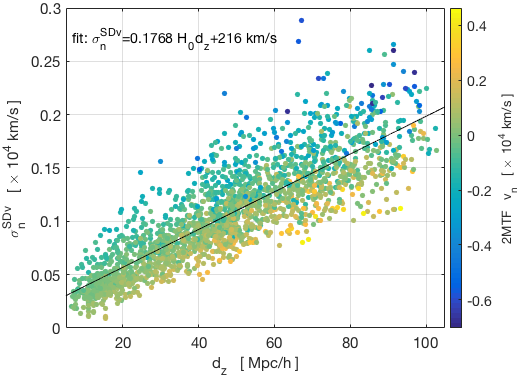}
 \caption{The linear fit between $\sigma_n^{SDv}$ and $d_z$ for 2MTF. The color indicates the PVs. The best fitting line is shown in black.}
\label{errvn}
\end{figure}

In an earlier 6dFGSv study, \cite{2016MNRAS.455..386S} calculate the measurement errors of the PVs by first converting the probability distribution function for $\eta$ for each galaxy (which is assumed to be Gaussian) to the PDF of velocities using 
\be\label{traPv}
\begin{split}
P(v)=&P(\eta)\frac{d\eta}{dv} =P(\eta)\frac{d\eta}{dd_h}\frac{dd_h}{dz_h}\frac{dz_h}{dv}\\ 
=&\frac{(1+z_h)^2}{(1+z)}\frac{P(\eta)}{\ln(10)H_0d_hE(z)}.
\end{split}
\ee
From this they define the standard deviation of the peculiar velocity as
\be\label{ssA3}
\sigma_n^{SDv}=\sqrt{\int(v-\bar{v})^2P(v)dv}=\sqrt{\int v^2P(v)dv-\bar{v}^2} 
\ee
where $\bar{v}$ is the velocity corresponding to the mean value of $\eta$. Finally, to remove the dependence of the uncertainty on the velocity itself, they fit the relationship between $\sigma_n^{SDv}$ and $d_{z}$ for each galaxy, finding
\be\label{appdcz1}
\sigma_n({\rm 6dFGSv})=0.324H_0d_z
\ee
which is then used as the error for each galaxy in the $d$MLE estimator. When applying the $d$MLE estimator to the 2MTF data, we follow the same procedure. Fig.~\ref{errvn} shows the inferred comoving distance of each 2MTF galaxy against the standard deviation in its peculiar velocity. Fitting this data gives
\be\label{appdcz2}
\sigma_n({\rm 2MTF})=0.177H_0d_z.
\ee

\section{Velocity dispersion errors in the 6\lowercase{d}FGS\lowercase{v} mocks}\label{AP2}

\begin{figure}  
 \includegraphics[width=\columnwidth]{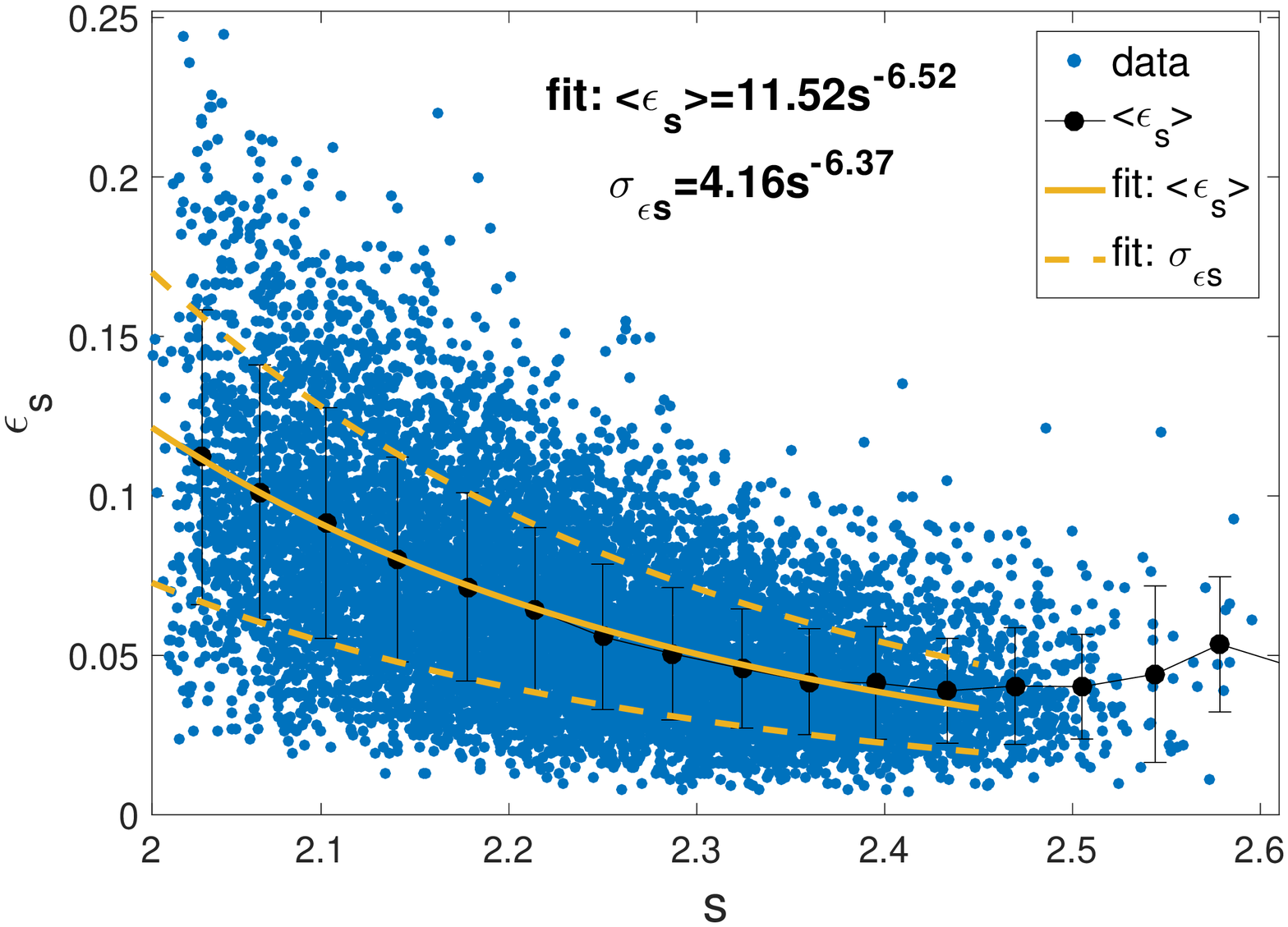}
 \caption{The relation between $s$ and $(\epsilon_s,~\sigma_{\epsilon_s})$ for 6dFGSv. Since the data is noisy for $s>2.45$, the fitting is constrained to the interval $s\in[2, 2.45]$.}
\label{es}
\end{figure}

To calculate the errors of the velocity dispersion, i.e $\epsilon_s$ for the 6dFGSv mocks, we fit the relationship between the error and the velocity dispersion itself from the 6dFGSv data, as shown in Fig.~\ref{es}. We calculate the mean $\langle \epsilon_{s} \rangle$, and standard deviation $\sigma_{\epsilon_{s}}$ of the error in 17 bins in $s$ and fit both of these with a power law, finding
\be
\langle \epsilon_{s} \rangle=11.52 s^{-6.52}~,~~\sigma_{\epsilon_s}=4.16 s^{-6.37} .
\ee
Given these fits, we generate $\epsilon_s$ for each mock galaxy under the assumption that the scatter in $\epsilon_{s}$ is Gaussian, according to:
\be
P(\epsilon_s)=\frac{1}{\sqrt{2\pi\sigma_{\epsilon_s}^2}} \exp\left(-\frac{(\epsilon_s-<\epsilon_s>)^2}{2\sigma_{\epsilon_s}^2}\right) .
\ee

\section{Bias in $B_z$ for 6\lowercase{d}FGS\lowercase{v}}\label{AP3}

In Section~\ref{sec:mocktests6dFGSv} we identified a bias in the $z$-component of bulk flow measured in our 6dFGSv mock galaxy catalogues that is also evident (although to a lesser degree) in the combined mocks, and which will affect the real data in a similar manner.

\begin{table*}   \small
\caption{The $\delta B_z$-revised bulk flow measurement in equatorial coordinates.}
\begin{tabular}{|c|c|c|c|c|c|c|c|}
\hline
\hline
\multicolumn{8}{|c|}{Equatorial Coordinates}\\
\hline

Data set   & $|\vec{B}|$ & $B_x$ &   $B_y$  &  $B_z$ 
& RA & Dec & Depth \\

&km s$^{-1}$ &   km s$^{-1}$  & km s$^{-1}$&km s$^{-1}$
& degree&degree &Mpc h$^{-1}$\\
\hline

6dFGSv& $201.2\pm32.9$  &$-198.3\pm33.6$    & $0.9\pm41.9$&  $-34.0\pm44.2$ &    $179.7\pm 12.0$&$-9.7 \pm 12.4$     & 75.6 \\

Combined& $311.9\pm26.7$  &$-225.6\pm20.8$    & $8.4\pm27.4$ &  $-215.2\pm29.9$ &    $177.9\pm 6.9$&$-43.6\pm  4.9  $ & 38.9 \\
\hline
          
\end{tabular}
 \label{bkflradecORI}
\end{table*}

\begin{table*}   \small
\caption{The $\delta B_z$-revised bulk flow measurement in Galactic coordinates.}
\begin{tabular}{|c|c|c|c|c|c|c|c|}
\hline
\hline
\multicolumn{8}{|c|}{Galactic Coordinates}\\
\hline
 
Data set   & $|\vec{B}|$ & $B_x$ &   $B_y$  &  $B_z$ 
& $\ell$ & $b$ & Depth \\

&km s$^{-1}$ &   km s$^{-1}$  & km s$^{-1}$&km s$^{-1}$
& degree&degree &Mpc h$^{-1}$\\
\hline

6dFGSv& $201.2\pm38.1$  &$26.5\pm43.9$    & $-123.8\pm43.4$ &  $156.4\pm38.7$ &    $282.1\pm 19.0$& $51.0\pm 10.3$     & 75.6 \\

Combined& $311.9\pm29.7$  &$109.1\pm30.3$    & $-275.9\pm30.6$ &  $95.9\pm31.9$ &    $291.6\pm 5.7$& $17.9\pm 5.3$    & 38.9 \\
\hline
          
\end{tabular}
 \label{bkflbORI}
\end{table*} 

\begin{figure}  
 \includegraphics[width=\columnwidth]{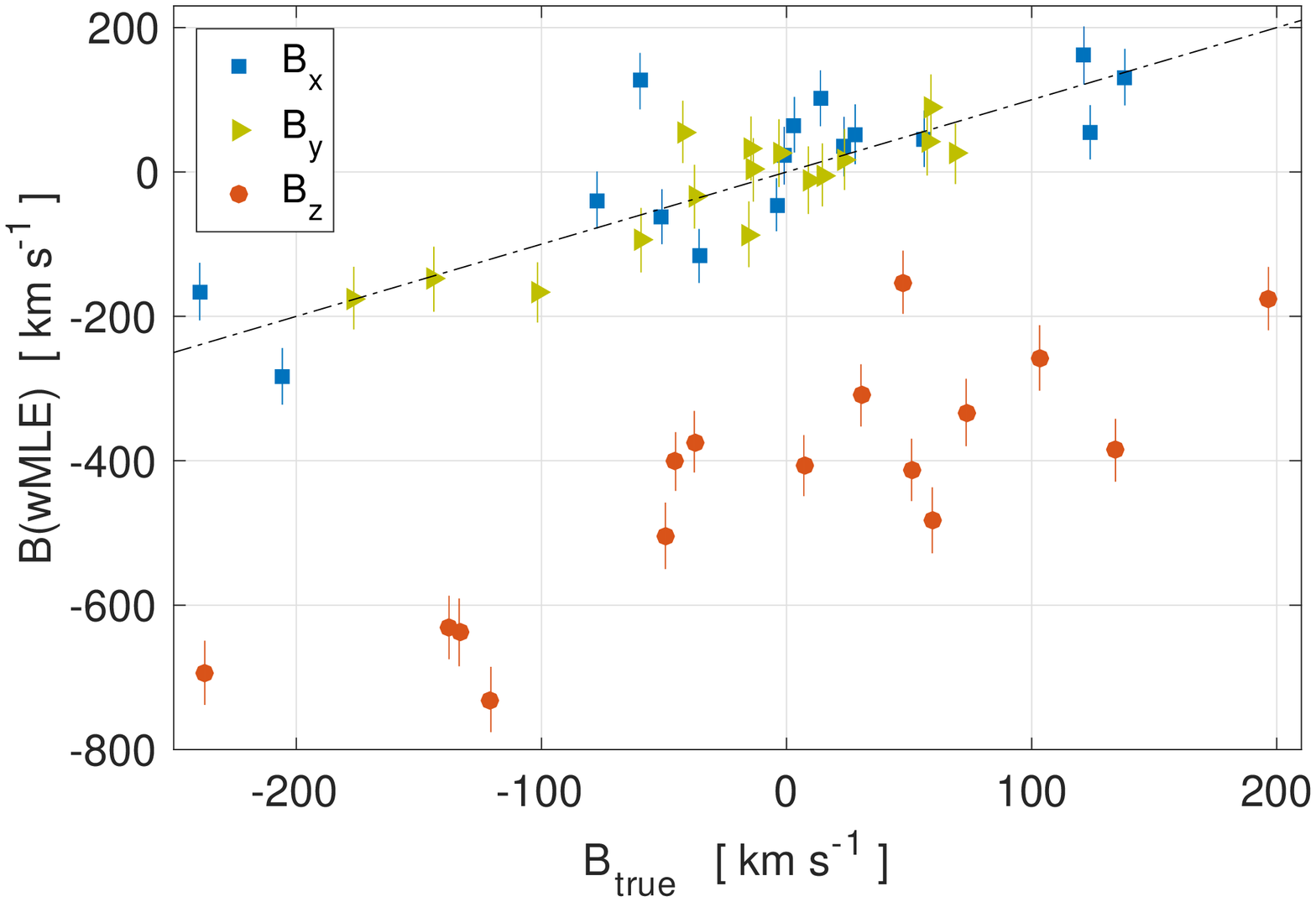}
 \caption{The measured bulk flow for the 6dFGSv mocks by using the $w$MLE.}
\label{esss}
\end{figure}

To test that this is not an estimator problem, we performed the same test using the $w$MLE method (Fig.~\ref{esss}) and using the minimum variance estimator as in \cite{2016MNRAS.455..386S}, but with velocities calculated from log-distance ratios using the estimator of \cite{2015MNRAS.450.1868W}. At a depth of $70h^{-1}\,\mathrm{Mpc}$, the MV results are: $B_x = -196\pm 43\,\mathrm{km\,s^{-1}}$, $B_y=-25\pm 51\, \mathrm{km\,s^{-1}}$, $B_z=-421\pm 66\,\mathrm{km\,s^{-1}}$ in equatorial coordinates. In both cases the results remain biased at approximately the same level, leading us to conclude this is not due to the estimator or the MLE method in general, but rather some aspect of the 6dFGSv survey and selection function. In the coordinate system we use, the $z$-direction corresponds to the vector directly along the southern pole and so an obvious assumption is that this bias is related to the hemispherical nature of the 6dFGSv data combined with a zero-point offset. 

Interestingly, the same test performed using the peculiar velocity estimator of \cite{2016MNRAS.455..386S} shows a smaller bias. However, we argue that the fact that this estimator does not preserve the true nature of the peculiar velocity distributions (compared to the \cite{2015MNRAS.450.1868W} estimator) means the bias is simply being hidden rather than removed, which is a more insidious problem. A more robust course of action is to use an estimator that conserves the correct probability distributions at the expense of highlighting the bias, and then correcting for this.

As the mocks have been shown to be biased and share the same characteristics as the real data, an alternative way to correct the bias in the $z$-component is to use the mean difference between the true and measured bulk 
flows $\delta B_z$, averaged over the 16 mocks. For the 6dFGSv mocks: $\delta B_z=405\pm 23\,\mathrm{km\,s^{-1}}$; for the combined mocks: $\delta B_{z}=132\pm 16\,\mathrm{kms^{-1}}$. The values of $\delta B_z$ can be added directly to the (biased) $B_{z}$ measurements from the data to produce de-biased results. These are presented in Tables~\ref{bkflradecORI} and \ref{bkflbORI}. When producing these de-biased results, we propagate the uncertainty in the correction factor into the uncertainty in $B_z$ using
the Jacobian in Section 4. Comparing Table~\ref{bkflradecORI} to Table~\ref{bkflradec}, we can see the $\delta B_z$-revised bulk flow velocities are very similar to the `$f_n$'-revised bulk flow velocities for 6dFGSv and the combined data set, which confirms that these corrections are robust.

\section{survey geometry of the Mocks}\label{AP34s}

\begin{figure}  
 \includegraphics[width=\columnwidth]{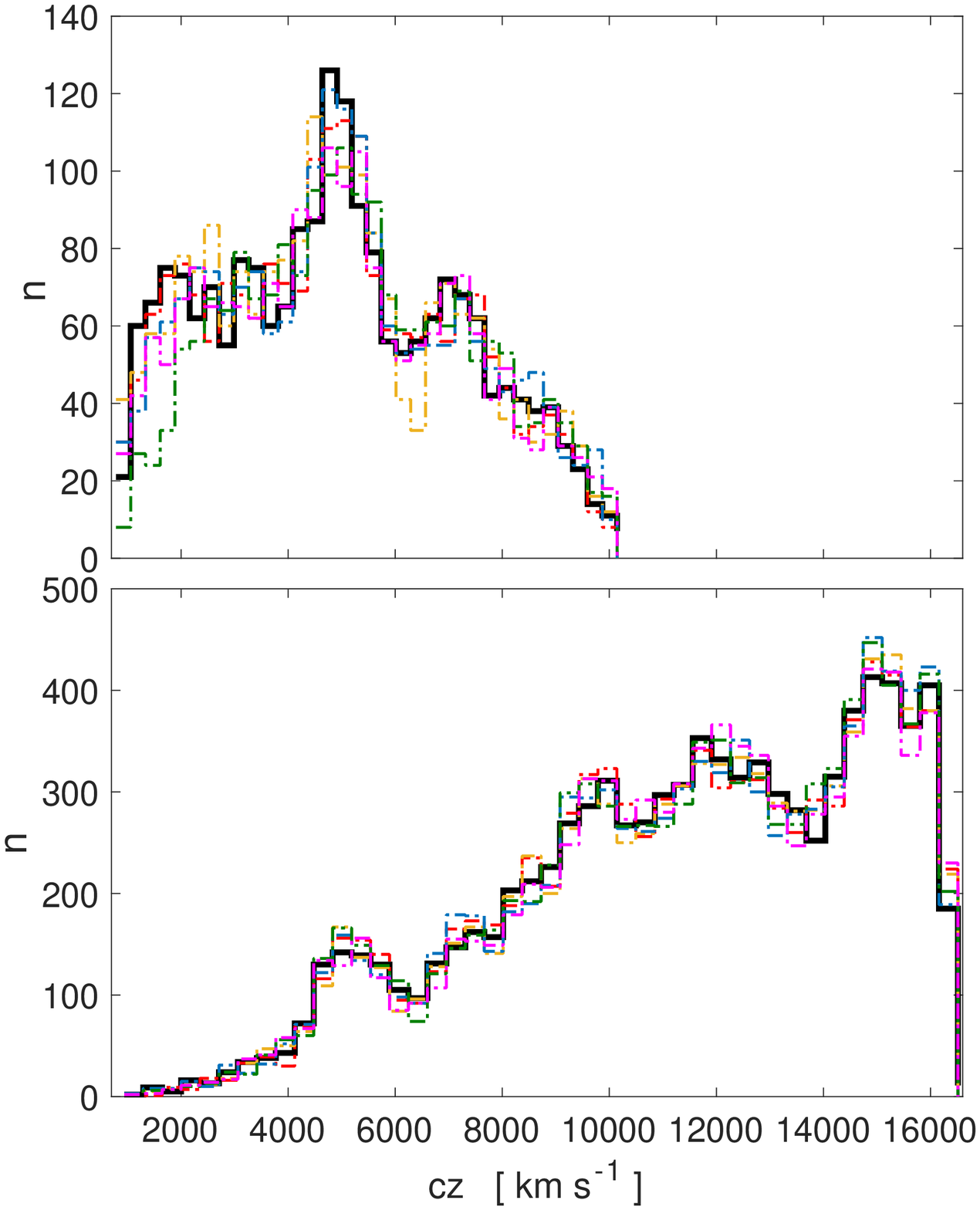}
 \caption{The distribution of $cz$ for the mocks and the 2MTF and 6dFGSv data. The upper panel is for 2MTF, with the black solid line representing the data, and the (coloured) dashed lines being the distribution of the mocks. Five example mocks are shown. The bottom panel is for 6dFGSv, with the black solid line representing the data, and the (coloured) dashed lines being the distribution of the mocks. Five example mocks are shown.}
 \label{ssD1}
\end{figure}

\begin{figure}  
 \includegraphics[width=\columnwidth]{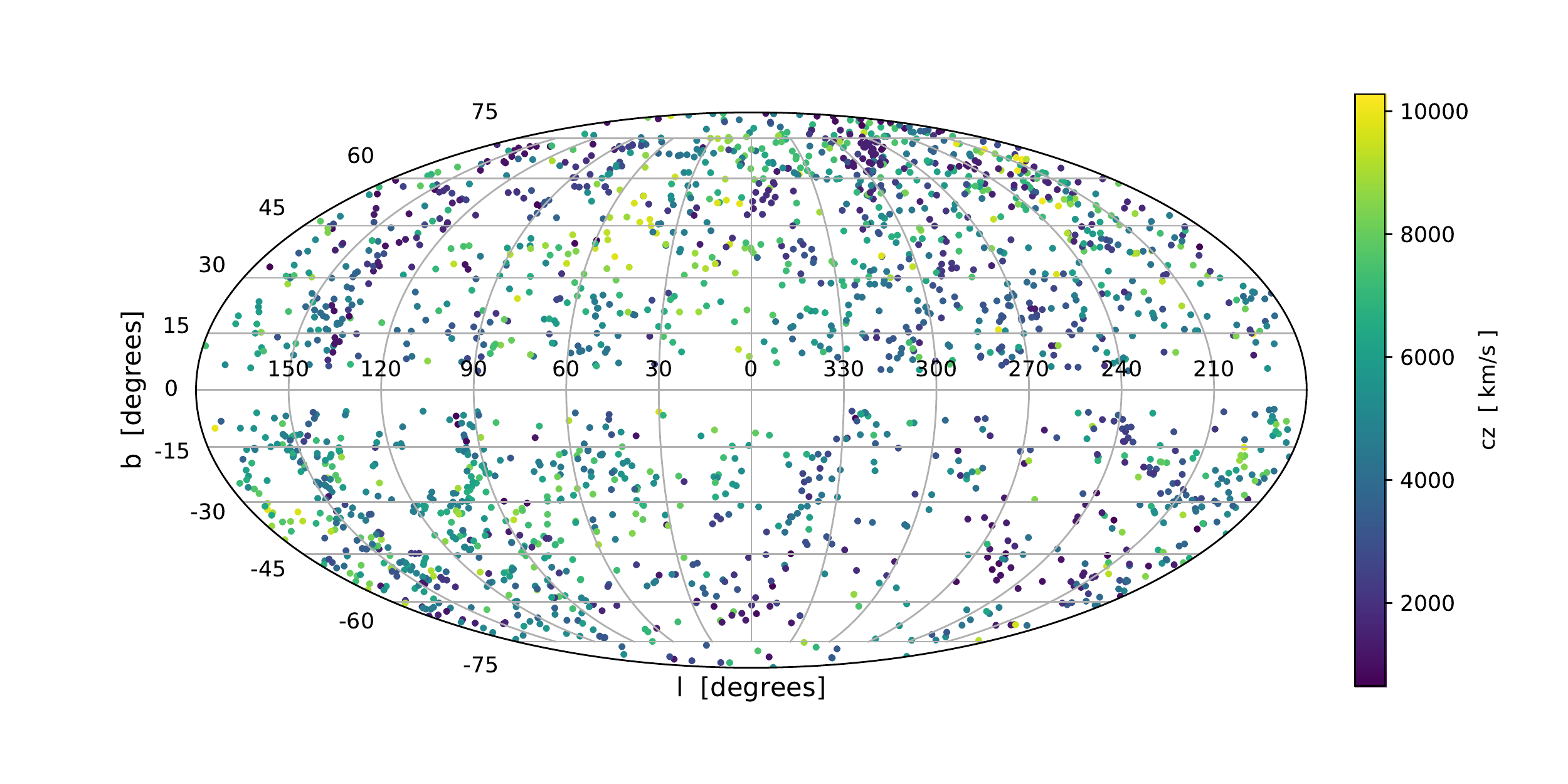}
  \includegraphics[width=\columnwidth]{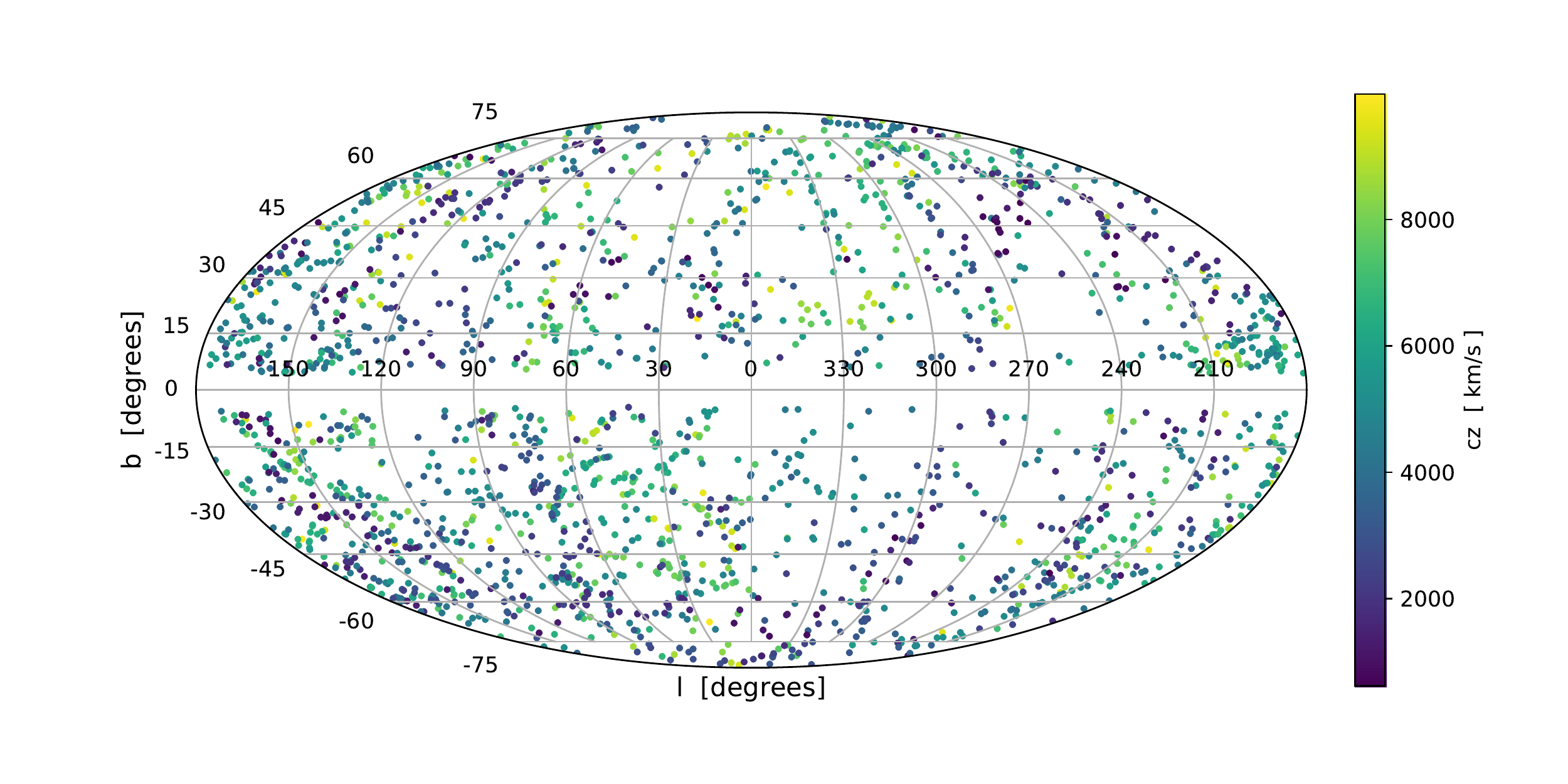}
 \caption{The sky coverage of the 2MTF data and a 2MTF mock. The upper panel is for the 2MTF data. The bottom panel is for an example 2MTF mock.}
  \label{ssD2}
\end{figure}

\begin{figure}  
 \includegraphics[width=\columnwidth]{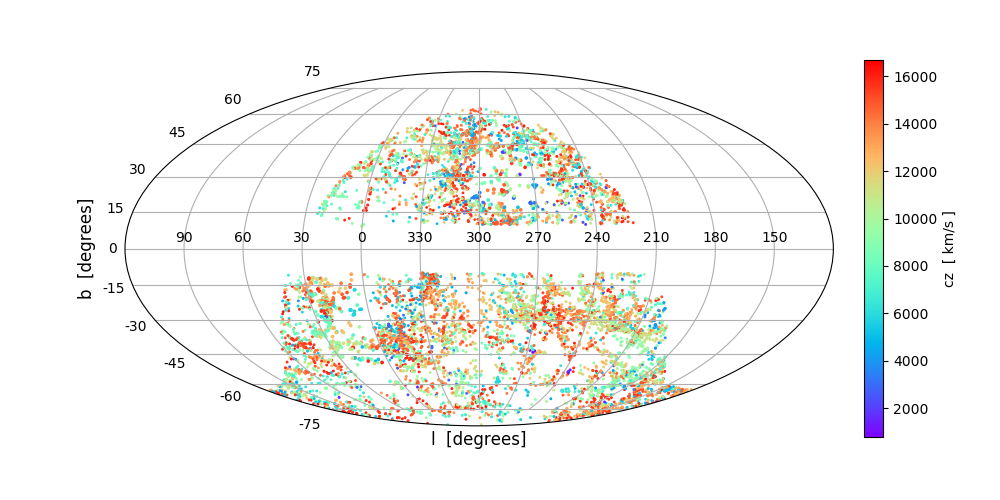}
  \includegraphics[width=\columnwidth]{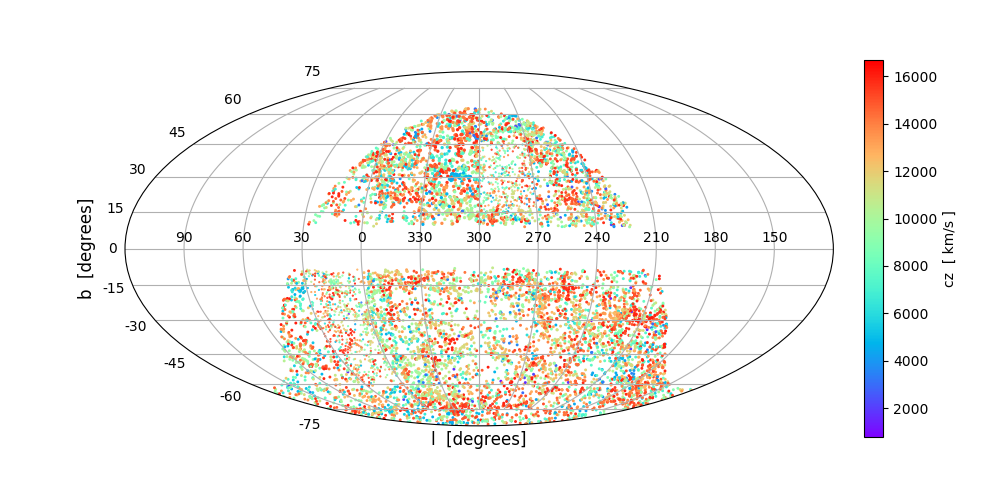}
 \caption{The sky coverage of the 6dFGSv data and a 6dFGSv mock. The upper panel is for the 6dFGSv data. The bottom panel is for an example 6dFGSv mock.}
  \label{ssD3}
\end{figure}

In Fig.~\ref{ssD1}, we plot the redshift distribution for five example 2MTF mocks alongside the real 2MTF data in the top panel, and five example 6dFGSv mocks alongside the real 6dFGSv data in the bottom panel. In Fig.~\ref{ssD2},
we plot the sky coverage of the 2MTF data in the top panel, and of an example 2MTF mock in the bottom panel. The sky coverage of the 6dFGSv data and an example 6dFGSv mock is shown in Fig.~\ref{ssD3}.

 \section{common galaxies in 2MTF and 6\lowercase{d}FGS\lowercase{v}}\label{AP33}

 \begin{table*} \tiny
\caption{The 43 common galaxies in 2MTF and 6dFGSv.}
\begin{tabular}{|c|c|c|c|c|c|c|c|}
\hline
\hline
\multicolumn{2}{|c|}{NAMES} & \multicolumn{3}{|c|}{6dFGSv PROPERTIES} & \multicolumn{3}{|c|}{2MTF PROPERTIES} \\
\hline
6dF Name & 2MASS Name &  $cz$ (km\,s$^{-1}$) & Dec (degree) & RA (degree) & $cz$ (km\,s$^{-1}$) & Dec (degree) & RA (degree)\\
\hline
g0011126-333443 & 2MASXJ00111259-3334428 & 7583 & -33.5786 & 2.80245 &  7587 & -33.5786 & 2.8025\\
\hline
g0022339-082911 & 2MASXJ00223386-0829109 & 5356 & -8.4864 & 5.6412 &  5335 & -8.4864 & 5.6411\\
\hline
g0047464-095006 & 2MASXJ00474641-0950063 & 5402 & -9.8350 & 11.94345 &  5430 & -9.8351 & 11.9434\\
\hline
g0121178-224802 & 2MASXJ01211776-2248024  & 5696 & -22.8007 & 20.32395 &  5636 & -22.8007 & 20.3240\\
\hline
g0122238-005231 & 2MASXJ01222375-0052308  & 7783 & -0.8752 & 20.59905 &  7823 & -0.8752 & 20.5990\\
\hline              
g0228201-315252  & 2MASXJ02282010-3152518 & 4466 & -31.8810 & 37.08375 & 4406 & -31.8811 & 37.0838 \\
\hline                          
g0230428-025620  &  2MASXJ02304283-0256204 & 5490 & -2.9390 & 37.67850 & 5498 & -2.9390 & 37.6785\\
\hline              
g0237587-015039  & 2MASXJ02375871-0150390 & 8346 & -1.8442 & 39.49470 & 8293 & -1.8442 & 39.4946\\
\hline            
g0247068-025821  & 2MASXJ02470675-0258213 &  7370 & -2.9726 & 41.77815 & 7358 & -2.9726 & 41.7781\\
\hline           
g0325017-054445  & 2MASXJ03250169-0544452 & 5536 & -5.7459 & 51.25710 & 5507 & -5.7459 & 51.2570 \\
\hline            
g0327292-213337  & 2MASXJ03272918-2133367 & 4123 & -21.5602 & 51.87165  & 4101 & -21.5602 & 51.8716\\
\hline           
g0348285-184508  & 2MASXJ03482846-1845082 & 9454 & -18.7523 & 57.11865 & 9435 & -18.7523 & 57.1186\\
\hline           
g0437363-044254  & 2MASXJ04373626-0442534 & 3696 & -4.7149 & 69.40110 & 3643 & -4.7148 & 69.4011\\
\hline            
g0452547-152047  & 2MASXJ04525466-1520472 & 5708 & -15.3465 & 73.22775  & 5685 & -15.3464 & 73.2278\\
\hline           
g0554452-150803  & 2MASXJ05544516-1508035 & 7453 & -15.1343 & 88.68825  & 7555 & -15.1343 & 88.6882\\
\hline           
g0557522-200505  & 2MASXJ05575221-2005047 & 3052 & -20.0846 & 89.46765 & 3050 & -20.0846 & 89.4675\\
\hline            
g0742549-711310  & 2MASXJ07425487-7113095  & 8509 & -71.2193 & 115.72860 & 8514 & -71.2193 & 115.7286\\
\hline            
g0843486-785658  & 2MASXJ08434862-7856577 & 5587 & -78.9494 & 130.95270 & 5597 & -78.9494 & 130.9526\\
\hline             
g0955567-134514  & 2MASXJ09555669-1345141 & 9672 & -13.7540 & 148.98615 & 9707 & -13.7539 & 148.9862\\
\hline            
g1018366-175857  & 2MASXJ10183654-1758571 & 3862 & -17.9825 & 154.65225 & 3887 & -17.9825 & 154.6523\\
\hline            
g1108222-475552  & 2MASXJ11082219-4755513 & 4687 & -47.9310 & 167.09249 & 4689 & -47.9309 & 167.0925\\
\hline            
g1111305-181722  & 2MASXJ11113045-1817219 &  4021 & -18.2895 & 167.87700 &  4115 & -18.2894 & 167.8769\\
\hline           
g1220373-184001 & 2MASXJ12203728-1840013 & 8592 & -18.6671 & 185.15535  & 8560 & -18.6670 & 185.1553\\
\hline            
g1257059-121620 & 2MASXJ12570592-1216194 & 6628 & -12.2721 & 194.27460 & 6674 & -12.2721 & 194.2747\\
\hline            
g1257229-153855 & 2MASXJ12572291-1538551 & 5756 & -15.6487 & 194.34540 & 5813 & -15.6486 & 194.3455\\
\hline             
g1258008-033716 & 2MASXJ12580082-0337161 & 5248 & -3.6211 & 194.50336 & 5277 & -3.6211 & 194.5034\\
\hline             
g1259341-210548 & 2MASXJ12593411-2105478 & 6678 & -21.0966 & 194.89215 & 6749 & -21.0966 & 194.8921\\
\hline             
g1353082-165737 & 2MASXJ13530820-1657371 & 6578 & -16.9604 & 208.28414 & 6601 & -16.9603 & 208.2842\\
\hline             
g1353097-304246 & 2MASXJ13530964-3042461 & 7160 & -30.7127 & 208.29030 & 7168 & -30.7128 & 208.2902\\
\hline              
g1353585-273724 & 2MASXJ13535843-2737234 & 5755 & -27.6232 & 208.49356 & 5803 & -27.6232 & 208.4935\\
\hline            
g1422254-342155 &  2MASXJ14222537-3421555 & 4218 & -34.3654 & 215.60580 & 4234 & -34.3654 & 215.6057\\
\hline            
g1506246-095426 &  2MASXJ15062462-0954258 & 7422 & -9.9072 & 226.60260 & 7487 & -9.9072 & 226.6026\\
\hline            
g1513457-141611 & 2MASXJ15134569-1416112 & 2136 & -14.2698 & 228.44055 & 2184 & -14.2698 & 228.4404\\
\hline             
g1537127-051957 & 2MASXJ15371268-0519572 & 8156 & -5.3326 & 234.30285 & 8110 & -5.3326 & 234.3028\\
\hline            
g1616036-223731 & 2MASXJ16160361-2237314 & 7744 & -22.6254 & 244.01505 & 7728 & -22.6254 & 244.0150\\
\hline             
g1852552-591520 & 2MASXJ18525522-5915196 & 3556 & -59.2554 & 283.23014 & 3603 & -59.2554 & 283.2301 \\
\hline           
g2000035-320505 & 2MASXJ20000350-3205052  & 5811 & -32.0848 & 300.01471 & 5755 & -32.0848 & 300.01460\\
\hline             
g2001027-170309 & 2MASXJ20010273-1703088 & 7601 & -17.0525 & 300.26130 & 7588 & -17.0524 & 300.2614\\
\hline            
g2017222-531710 & 2MASXJ20172214-5317101 & 4344 & -53.2862 & 304.34235 & 4341 & -53.2861 & 304.3423\\
\hline            
g2102420-171633 & 2MASXJ21024203-1716324 & 8590 & -17.2757 & 315.67514 & 8559 & -17.2757 & 315.6751 \\
\hline            
g2130142-080401 & 2MASXJ21301423-0804011 & 8521 & -8.0670 & 322.55925 & 8501 & -8.0670 & 322.5593\\
\hline            
g2136487-224820 & 2MASXJ21364872-2248195  & 8101 & -22.8054 & 324.20294 & 8130 & -22.8054 & 324.2030 \\
\hline            
g2205270-003201 & 2MASXJ22052701-0032010  & 9090 & -0.5336 & 331.36259 & 8959 & -0.5336 & 331.3625 \\
\hline          
\end{tabular}
 \label{tbbb}
\end{table*}


\label{lastpage}
\end{document}